\documentclass[sigconf, screen, 10pt]{acmart}

\usepackage{amsmath}
\usepackage{amsfonts}
\usepackage[T1]{fontenc}
\usepackage[utf8]{inputenc} 
\usepackage{booktabs}
\usepackage[english]{babel}
\usepackage{blindtext}
\usepackage{color}
\usepackage{soul}
\usepackage{xspace}
\usepackage[symbol]{footmisc}
\usepackage{enumitem}
\usepackage{float}
\usepackage{tikz}
\usetikzlibrary{arrows.meta}
\usetikzlibrary{automata,positioning}
\usetikzlibrary{calc}
\usetikzlibrary{arrows.meta,arrows,backgrounds,patterns,positioning,shadings, decorations.markings, shapes.multipart, shapes}
\usepackage{subcaption}

\usepackage[ruled,vlined,linesnumbered]{algorithm2e}
\usepackage{xurl}
\usepackage{siunitx}
\usepackage{graphicx}
\usepackage{multirow}
\usepackage{listings}
\usepackage{pgfplots, pgfplotstable}
\pgfplotsset{compat=newest}
\usepackage{arydshln}
\usepackage{wasysym}
\usepackage[framemethod=TikZ]{mdframed}
\mdfsetup{
   roundcorner=2.4pt}
\usepackage{cprotect}

\newcommand{\VARIADIC}[6]{%
  \expandafter\newcommand\csname Gobble#1Arg\endcsname[2]{%
    \csname Check#1Arg\endcsname{##1#4##2}%
  }%
  \expandafter\newcommand\csname Check#1Arg\endcsname[1]{%
    \csname @ifnextchar\endcsname\bgroup{\csname Gobble#1Arg\endcsname{##1}}{#2{##1#5}#6}%
  }%
  \expandafter\newcommand\csname #1\endcsname[1]{%
    \csname Check#1Arg\endcsname{#3##1}%
  }%
}
\newcommand{\delidot}{.}
\newcommand{\deli}{\mathbin{\mkern2mu\resizebox{3pt}{!}{\raisebox{1pt}{$\circ$}}\mkern2mu}}

\VARIADIC{lDN} {\begin{tt}} {} {\delidot} {\delidot}       {\end{tt}\xspace}
\VARIADIC{sDN} {\begin{tt}} {} {\delidot} {}               {\end{tt}\xspace}
\VARIADIC{DN}  {\begin{tt}} {} {\deli}    {}               {\end{tt}}
\VARIADIC{eDN} {\begin{tt}} {} {\deli}    {\deli \epsilon} {\end{tt}\xspace}

\usepackage[capitalise,nameinlink]{cleveref}
\crefname{algorithm}{Algorithm}{Algorithms}
\Crefname{algorithm}{Algorithm}{Algorithms}
\crefname{section}{\S\!}{\S\S\!}
\Crefname{section}{Section}{Sections}
\crefname{figure}{Figure}{Figures}
\Crefname{figure}{Figure}{Figures}
\crefname{equation}{Equation}{Equations}
\Crefname{equation}{Equation}{Equations}
\crefname{listing}{Listing}{Listings}
\Crefname{listing}{Listing}{Listings}
\crefname{defn}{definition}{definitions}

\newcommand{\OMIT}[1]{}

\newcommand{\EG}{\emph{e.g.}\xspace}
\newcommand{\IE}{\emph{i.e.}\xspace}
\newcommand{\ETC}{\emph{etc.}\xspace}

\newcommand{\tool}{\textsc{Eywa}\xspace}

\newcommand{\para}[1]{\textbf{#1.}\xspace}

\renewcommand\footnotetextcopyrightpermission[1]{}
\setcopyright{none}
\definecolor{darkgreen}{RGB}{20, 33, 61}
\definecolor{darkblue}{RGB}{20, 33, 61}
\definecolor{darkred}{rgb}{0.65,0,0}

\lstdefinestyle{CStyle}{
    language=C,
    basicstyle=\ttfamily\small,
    commentstyle=\color{green!40!black},
    keywordstyle=\bfseries\color{black},
    numbers=none,
    numberstyle=\tiny,
    numbersep=5pt,
    frame=single,
    showstringspaces=false,
    numberstyle=\scriptsize\color{gray},
    xleftmargin=0pt,
    xleftmargin=.75em,
    framexleftmargin=.5em,
    upquote=true,
    showspaces=false,
    showtabs=false,
    tabsize=2,
    columns=flexible,
    frame=none,
    literate={€}{{}}1,
    morekeywords={printf, scanf, int, void, if, else, while, for, return, bool, uint32_t, strcpy, strcmp}
}
\lstdefinestyle{PythonStyle}{
    language=Python,
    basicstyle=\ttfamily\small,
    commentstyle=\color{green!40!black},
    keywordstyle=\bfseries\color{blue},
    stringstyle=\color{darkred},
    numbers=none,
    numberstyle=\tiny,
    numbersep=5pt,
    frame=single,
    showstringspaces=false,
    numberstyle=\scriptsize\color{gray},
    xleftmargin=0pt,
    xleftmargin=.75em,
    framexleftmargin=.5em,
    upquote=true,
    showspaces=false,
    showtabs=false,
    tabsize=2,
    columns=flexible,
    frame=none,
    literate={€}{{}}1
}
\lstdefinestyle{TextStyle}{
    language={},
    basicstyle=\ttfamily\small,
    keywordstyle=\color{black},
    numbers=none,
    numberstyle=\tiny,
    numbersep=5pt,
    frame=single,
    showstringspaces=false,
    numberstyle=\scriptsize\color{gray},
    xleftmargin=0pt,
    xleftmargin=.75em,
    framexleftmargin=.5em,
    upquote=true,
    showspaces=false,
    showtabs=false,
    tabsize=2,
    columns=flexible,
    frame=none,
    literate={€}{{}}1
}
\newcommand{\ccode}[1]{\lstinline[mathescape, style=CStyle]{#1}}
\newcommand{\pcode}[1]{\lstinline[mathescape, style=PythonStyle]{#1}}

\definecolor{forestgreen}{rgb}{0.13, 0.65, 0.13}

\newcommand*\smallcircled[1]{\tikz[baseline=(char.center)]{
       \node[inner sep=2pt, draw, solid, circle, fill=blue!10] (char)
        {#1};}}
\newcommand*\smallcircledinline[1]{\tikz[baseline=(char.center)]{
       \node[shape=circle,draw,black,solid,minimum size=3.1mm, inner sep=0pt, fill=blue!10,text=black] (char)tube bdsm 
        {\footnotesize #1};}}
\newcommand{\inlinecircle}[1]{\raisebox{2.5pt}{\smallcircledinline{#1}}}

\newcommand{\ns}{\ensuremath{\mathtt{NS}}\xspace}
\newcommand{\soa}{\ensuremath{\mathtt{SOA}}\xspace}

\newcommand{\dname}{\ensuremath{\mathtt{DNAME}}\xspace}
\newcommand{\cname}{\ensuremath{\mathtt{CNAME}}\xspace}

\newcommand{\tuple}[2]{\left\langle #1,\ #2 \right\rangle}

\newcommand{\node}{node}

\makeatletter
\newcommand{\customlabel}[3]{
   \protected@write \@auxout {}{\string \newlabel {#1}{{#2}{\thepage}{#2}{#1}{}} }
   \hypertarget{#1}{#3}
}

\makeatother

\newsavebox\codea
\newsavebox\codeb
\newsavebox\codee
\newsavebox\codef
\newsavebox\codeg
\newsavebox\codeh
\newsavebox\codei
\newsavebox\codej
\newsavebox\codek
\newsavebox\codel

\acmConference[Submitted for review]{}
\acmYear{2025}
\begin{document}

\title{\tool: Automating Model Based Testing using LLMs}


\author{Rajdeep Mondal*}
\affiliation{\institution{UCLA}\city{}\country{}}

\author{Rathin Singha*}
\affiliation{\institution{UCLA}\city{}\country{}}

\author{Todd Millstein}
\affiliation{\institution{UCLA}\city{}\country{}}

\author{George Varghese}
\affiliation{\institution{UCLA}\city{}\country{}}

\author{Ryan Beckett}
\affiliation{\institution{Microsoft Research}\city{}\country{}}

\author{\hspace{-0.135em}Siva Kesava Reddy Kakarla}
\affiliation{\institution{Microsoft Research}\city{}\country{}}

\begin{abstract}
Model-based testing (MBT), whereby a model of the system under test is analyzed 
to generate high-coverage test cases, has been 
used to test 
protocol implementations. 
A key barrier to the use of MBT is the need for users to understand protocol RFCs in detail to create a compliant model. 
Our new approach to MBT uses LLMs to automatically build rich models of intended protocol behavior 
from knowledge embedded in RFCs, blogs, 
and other natural language sources. Our approach addresses key challenges with using LLMs, including hallucinations and their inability to monolithically generate complex protocol models. 
We realize our approach through a novel protocol testing framework \tool,  
and demonstrate 
its effectiveness through extensive case studies of DNS and BGP 
and a smaller study of SMTP. Despite 
minimal user effort, applying 
\tool enabled the discovery of
32 unique bugs across widely used DNS, BGP, and SMTP implementations, 15 of which were previously undiscovered despite extensive prior testing with manually crafted models. 
\end{abstract}

\maketitle

\def\thefootnote{*}\footnotetext{These authors contributed equally to this work}\def\thefootnote{\arabic{footnote}}

%
%
%
%

\section{Introduction}
\label{sec:introduction}

\begin{lrbox}{\codea}
\begin{lstlisting}[style=PythonStyle]
# Define the data types.
domain_name = eywa.String(maxsize=5)
record_type = eywa.Enum("RecordType", ["A", "AAAA", "NS", "TXT", "CNAME", "DNAME", "SOA"])
record = eywa.Struct("RR", rtyp=record_type, name=domain_name, rdat=eywa.String(3))
# Define the module arguments.
query = eywa.Arg("query", domain_name, "A DNS query domain name.")
record = eywa.Arg("record", record, "A DNS record.")
result = eywa.Arg("result", eywa.Bool(), "If the DNS record matches the query.")
# Define 3 modules to validate the query and implement the record matching logic.
valid_query = eywa.RegexModule("[a-z*](\.[a-z*])*", query)
da = eywa.FuncModule("dname_applies", "If a DNAME record matches a query.", [query, record, result])
ra = eywa.FuncModule("record_applies", "If a DNS record matches a query.", [query, record, result])
# Create the dependency graph to connect the modules.
g = eywa.DependencyGraph(); g.Pipe(valid_query, ra); g.CallEdge(ra, [da]);
# Synthesize the end-to-end model and generate test inputs.
model = g.Synthesize(main=ra)
inputs = model.generate_tests(timeout="300s")    
\end{lstlisting}
\end{lrbox}

\begin{lrbox}{\codeb}
\begin{lstlisting}[style=CStyle]
// Pre-defined modules implemented by Eywa.
bool valid_query(char* query) { ... }
// Modules implemented by the LLM.
bool dname_applies(char* query, Record record) { ... }
bool record_applies(char* query, Record record) { ... }
// Symbolic test harness for KLEE.
int main(){
    char x0[6]; Record x1;
    ...
    bool result_tmp; bool x3;
    klee_make_symbolic(x3, sizeof(x3), "x3");
    bool bad_input; bool x4;
    klee_make_symbolic(x21, sizeof(x4), "x4");
    if (valid_query(x0)) {
        bad_input = false;
        result_tmp = record_applies(x0, x1);
    }
    else{
        bad_input = true;
        result_tmp = false;
    }
    klee_assume(result_tmp == x3); 
    klee_assume(bad_input == x4);
}
\end{lstlisting}
\end{lrbox}



Networked systems depend on the correct and performant operation of 
hundreds of protocols from the physical to the application layer. Despite 
standardization, protocol hardware and software implementations still frequently suffer from bugs due to coding mistakes, misinterpretations of specifications, unsound optimizations, unforeseen corner cases, and poor data structure choices. Bugs in protocol implementations have led to security vulnerabilities~\cite{scale, bug-ipnet, bug-pop3, bug-ftp1, bug-ftp2, bug-icmp, bug-cisco}, 
outages~\cite{outage-bgp, outage-akamai, bug-juniper, outage-rogers, outage-medical, bug-ios, outage-ripe}, and 
performance problems~\cite{congestion-avoidance}. For example, in 2022 a bug in Akamai's DNS software caused a global outage for nearly 30,000 sites~\cite{outage-akamai}. Recently, mishandling of a corrupted BGP attribute in router software caused invalid routes to spread 
before shutting down BGP sessions in remote networks~\cite{outage-bgp}.



To test the correctness of protocol implementations, model-based testing (MBT) has emerged as a highly effective family of techniques. Rather than writing test cases manually, in MBT the user formulates a simplified model of a protocol -- for instance as a logical specification, abstract state machine, or a reference implementation -- and then uses existing program analysis techniques such as symbolic execution~\cite{klee, dart} to enumerate test cases from this model that cover a wide variety of protocol behaviors. 
Researchers have used MBT to find numerous bugs in implementations of QUIC~\cite{quic, quic2}, BGP~\cite{messi}, DNS~\cite{scale, dns_maude}, and various LTE~\cite{tfuzz} protocols.

Since handwritten protocol models are typically orders of magnitude simpler than the 
implementation's source code, symbolic execution scales well on these models and ameliorates the infamous path explosion problem~\cite{klee} that arises 
when analyzing source code directly. 
MBT also applies to testing proprietary protocol implementations where the source code is unavailable but the 
standards are available. However, a key bottleneck to using MBT is 
that users must carefully build a model of the protocol under test. This requires 
reading multiple RFCs to understand 
the detailed protocol behavior, and 
to translate that understanding into either a 
formal model or a 
reference implementation. Hence, applying MBT to any new protocol incurs a significant up-front cost before 
testing, 
thereby limiting its applicability in practice.

In this work, we ask: 
\textit{``can we automatically derive protocol models for testing using LLMs?''} We answer this question affirmatively by exploiting 
recent advances in large language models (LLMs) 
to use their knowledge base for automated testing. For instance, as part of their training data, LLMs have 
ingested substantial protocol knowledge from RFCs, standards, networking forums \& blogs, as well as other online resources and documents. However, using LLMs to test network protocols in practice requires overcoming several non-trivial challenges.

\begin{enumerate}[leftmargin=*]
    \setlength{\itemsep}{2pt}
    \item \textbf{LLM errors:} LLMs make mistakes and can easily hallucinate to produce flawed protocol models.
    \item \textbf{Valid inputs:} Protocol inputs can have complex requirements and data dependencies. For instance, creating a valid TCP header, a DNS zone, or a BGP route to test is non-trivial, and invalid inputs can render a test useless.
    \item \textbf{Large models:} Many protocols are highly complex and comprise multiple components. Asking an LLM to generate an end-to-end protocol model in a single shot is impractical and increases the likelihood of model errors.
    \item \textbf{Protocol state}: While 
    DNS is stateless, BGP depends on 
    earlier routes, and protocols like SMTP and TCP use state machines that require the right sequence of inputs to drive an implementation to ``deeper'' states.
\end{enumerate}

In this paper, we describe an approach that uses LLMs to automatically derive protocol models for MBT while addressing each of the challenges described above. 
%
In our approach, the user provides (1) a short description of which protocol component they wish to test, (2) its input and output types, and (3) a graph that specifies how the different components (or ``modules'') should be composed together. Our testing framework then automatically generates a model for each protocol module, and it uses the given graph to stitch these modules together to produce the end-to-end protocol model. The testing framework then performs symbolic execution on this model to systematically enumerate test cases for the protocol that the user can then run against real protocol implementations.


Our approach effectively resolves the challenges 
identified earlier. First,
because LLMs can produce flawed models (\textit{challenge \#1}), rather than relying on the model output 
we use \emph{differential testing} on multiple implementations; if the tests produce different outputs in different protocol implementations, often some implementation is buggy. Differential testing is particularly natural for protocol testing, where 
we often have many implementations that must interoperate.


Second, by decomposing complex protocols like BGP into smaller modules that are composed through an explicit dependency graph, the LLM can scale to construct larger protocol models (\textit{challenge \#3}). When constructing each protocol module, the LLM has access to the type signatures and descriptions of depended upon modules, but not their implementation details. Enforcing validity constraints on protocol inputs (\textit{challenge \#2}) is achieved through a simple form of sequential composition, where the user defines one module to enforce the validity constraints and pushes its output into a second module that assumes the inputs are valid.

Finally, we handle stateful protocols (\textit{challenge \#4}) through a separate invocation of the LLM to generate a state graph of the protocol. For stateful protocols, it is not enough to create the model and generate tests as the test cases are state-input pairs. In order to run these tests, we also need to drive the protocol implementation to the state required by the test cases. We use this generated state graph to search for inputs that lead to particular protocol states and then use that input sequence to drive the protocol implementation to those states as needed for different tests.


We have instantiated our approach in a tool called \tool, which is implemented as a Python library. Given short descriptions of protocol modules and a dependency graph, \tool automatically constructs a protocol model through multiple modular invocations of an LLM to produce executable C code. \tool simultaneously compiles a \emph{symbolic test harness} to integrate this model with the Klee~\cite{klee} symbolic execution engine. By invoking Klee on the end-to-end protocol model, \tool obtains a high-coverage set of test cases and translates them back to Python for the library user. The user can then apply these tests to real protocol implementations.

We evaluate the effectiveness of \tool through an extensive case study of the DNS and BGP protocols, and a smaller study of SMTP as a canonical stateful protocol. Compared to prior model-based testing approaches for DNS and BGP that required months of effort to craft models that accurately encode the RFC intent~\cite{scale, messi}, \tool constructed similar models with minimal user input. Using \tool, we revealed \textbf{32} unique bugs across popular DNS, BGP, and SMTP implementations. Of these bugs, \textbf{15} were were previously unknown despite the extensive testing by prior work.

\para{Contributions}
To summarize, our contributions are:
\begin{itemize}
    \item A new approach to protocol testing that combines LLMs with model-based testing and differential testing.
    \item \tool, a protocol testing library based on our approach that provides novel abstractions for defining and composing protocol modules.
    \item Techniques to compile \tool specifications by generating specialized LLM prompts and combining the resulting models with a \emph{symbolic test harness}.
    \item Case studies that showcase \tool's utility that found 32 unique bugs across popular DNS, BGP and SMTP implementations, including 15 previously undiscovered bugs.
    \item A small case study of SMTP that shows how to handle stateful protocols by using the LLM separately to generate a state machine of the protocol as a graph, and using graph traversal to automatically drive an implementation to desired states.
\end{itemize}

\textbf{Ethics:} This work raises no ethical concerns.

%
%
%
%

\section{Motivation and Overview}
\label{sec:overview}


To motivate \tool, we demonstrate its use to automatically generate tests for Domain Name System (DNS) nameservers. We selected DNS because its analysis has been the target of prior model-based testers. SCALE~\cite{scale} was the first to apply MBT to the DNS protocol and found dozens of new correctness, performance, and security bugs in widely used DNS implementations such as BIND, Knot, PowerDNS, and CoreDNS. To achieve these results however, SCALE required extensive manual mathematical formulation of the DNS RFCs~\cite{groot} and later arduous manual construction of an executable model~\cite{scale}.
We give an overview of \tool in the context of DNS testing. Our later experiments show that \tool achieves results similar to SCALE, including finding 11 new DNS bugs that SCALE missed, all with minimal user effort. 

\begin{lrbox}{\codee}
\begin{lstlisting}[style=TextStyle]
klee --libc=uclibc
     --posix-runtime
     --max-time=300s
     --external-calls=all program.bc
\end{lstlisting}
\end{lrbox}

\begin{figure*}[ht!]
  \centering
  \begin{tikzpicture}
    \node[inner sep=3pt,fill=white!7] (background) at (0,0) {
      \begin{tikzpicture}
        \node[inner sep=2pt] (codea) at (-.05,0) {
          \begin{tikzpicture}
            \node[inner sep=5pt,rounded corners=.1cm,draw=gray,fill=gray!0,thick,label={below:(a) User input to \tool to generate a model for DNS lookup.}] (eywacode) at (0,0) { \usebox\codea };
          \end{tikzpicture}\label{fig:modelPython}};
        \node[inner sep=2pt,,label={below:(b) Generated implementation and test harness.}] (codeb) at (-3.6,-10) {
          \begin{tikzpicture}
            \node[inner sep=5pt,rounded corners=.1cm,draw=gray,fill=gray!0,thick] (ccode1) at (0,0) { \usebox\codeb };
          \end{tikzpicture}};
        \node[inner sep=2pt] (klee) at (5.5,-14.5) {
          \begin{tikzpicture}
            \node[inner sep=5pt,rounded corners=.1cm,draw=gray,fill=gray!0,thick,label={below:(c) Klee invocation.}] (ccode) at (7,-10) { \usebox\codee };
          \end{tikzpicture}};
        \node[label={below:(d) Differential testing setup.}] (docker) at (5.5, -10.6) { \includegraphics[width=1.4cm]{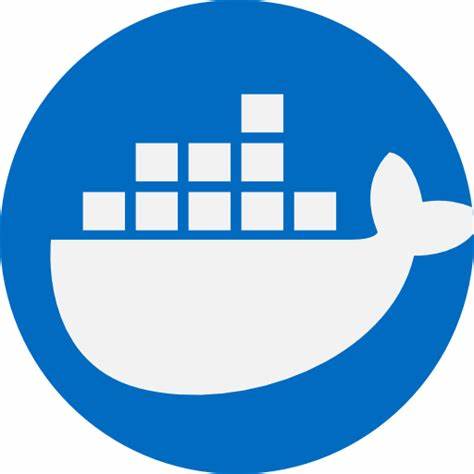} };
        \node[label={}] (ex1) at (4.5, -12.6) { Test inputs };
        \node[label={}] (server1) at (3.5, -7.8) { \includegraphics[width=1cm]{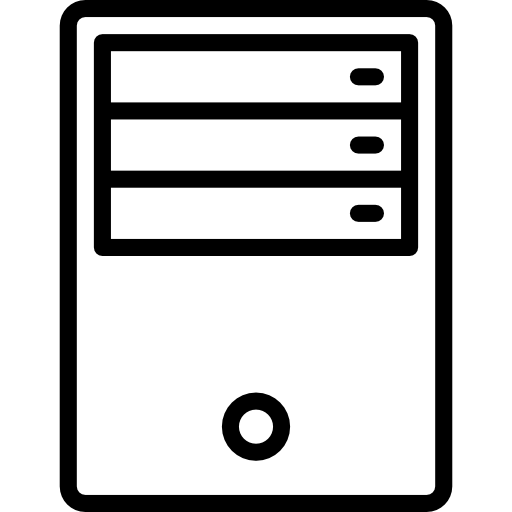} };
        \node[label={}] (server2) at (4.5, -7.8) { \includegraphics[width=1cm]{icons/server.png} };
        \node[label={below: (e) Protocol implementations.}] (server3) at (5.5, -7.8) { \includegraphics[width=1cm]{icons/server.png} };
        \node[label={}] (server4) at (6.5, -7.8) { \includegraphics[width=1cm]{icons/server.png} };
        \node[label={}] (server5) at (7.5, -7.8) { \includegraphics[width=1cm]{icons/server.png} };
        \node[label={}] (results) at (5.5, -5.6) { \includegraphics[width=1.2cm]{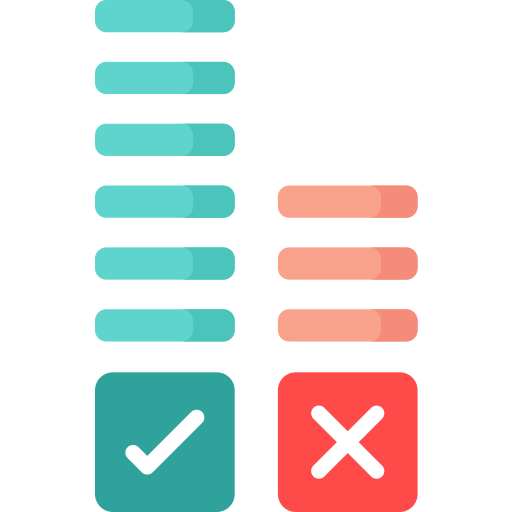} };
        \draw[solid,line width=.3mm,darkblue,rounded corners=.1cm,->] (5.5,-3.58) -- (5.5,-4.5) -- (2.5, -4.5) -- (2.5, -5.7) -- (.95,-5.7);
        \draw[solid,line width=.3mm,darkblue,rounded corners=.1cm,->](5.5,-3.58) -- (5.5,-4.5) -- (2.5, -4.5) -- (2.5, -10.5) -- (.95,-10.5);
        \draw[solid,line width=.3mm,darkblue,rounded corners=.1cm,->] (.95,-14.3) -- (2.5,-14.3);
        \draw[solid,line width=.3mm,darkblue,rounded corners=.1cm,->] (5.5,-13.2) -- ++(0, 1.3);
        \draw[solid,line width=.3mm,darkblue,rounded corners=.1cm,->] (5.5, -9.9) -- (5.5, -9);
        \draw[solid,line width=.3mm,darkblue,rounded corners=.1cm,->] (5.5, -9.9) -- (5.5, -9.6) -- (3.5,-9.6) -- (3.5,-9);
        \draw[solid,line width=.3mm,darkblue,rounded corners=.1cm,->] (5.5, -9.9) -- (5.5, -9.6) -- (4.5,-9.6) -- (4.5,-9);
        \draw[solid,line width=.3mm,darkblue,rounded corners=.1cm,->] (5.5, -9.9) -- (5.5, -9.6) -- (6.5,-9.6) -- (6.5,-9);
        \draw[solid,line width=.3mm,darkblue,rounded corners=.1cm,->] (5.5, -9.9) -- (5.5, -9.6) -- (7.5,-9.6) -- (7.5,-9);
        \draw[solid,line width=.3mm,darkblue,rounded corners=.1cm,->] (5.5, -7.2) -- (results.south);
        \draw[solid,line width=.3mm,darkblue,rounded corners=.1cm,->] (3.5, -7.2) -- (3.5, -6.8) -- (5.5,-6.8) -- (results.south);
        \draw[solid,line width=.3mm,darkblue,rounded corners=.1cm,->] (4.5, -7.2) -- (4.5, -6.8) -- (5.5,-6.8) -- (results.south);
        \draw[solid,line width=.3mm,darkblue,rounded corners=.1cm,->] (6.5, -7.2) -- (6.5, -6.8) -- (5.5,-6.8) -- (results.south);
        \draw[solid,line width=.3mm,darkblue,rounded corners=.1cm,->] (7.5, -7.2) -- (7.5, -6.8) -- (5.5,-6.8) -- (results.south);
        \node (label1) at (1.65,-5.2) {\customlabel{circle:1}{\protect\inlinecircle{$1$}}{\smallcircled{$1$}}};
        \node (label2) at (1.65,-10) {\customlabel{circle:2}{\protect\inlinecircle{$2$}}{\smallcircled{$2$}}};
        \node (label3) at (1.65,-13.8) {\customlabel{circle:3}{\protect\inlinecircle{$3$}}{\smallcircled{$3$}}};
        \node (label4) at (5.9,-12.6) {\customlabel{circle:4}{\protect\inlinecircle{$4$}}{\smallcircled{$4$}}};
      \end{tikzpicture}};
  \end{tikzpicture}
  \vspace{-1em}
  \caption{Example of how \tool generates tests. It uses the LLM to implement the module logic . It then compiles a symbolic test harness for Klee  and invokes Klee to perform symbolic execution  and generate test cases. Tests generated from the \tool runtime are then executed for every implementation.}
  \label{fig:overview-example}
\end{figure*}

\subsection{Testing the DNS with \tool}
\label{subsec:overview-dns}

To test a protocol, users must describe the parts of that protocol that they wish to model using \tool's Python library abstractions. Specifically, users define protocol-specific objects (\EG, state) and formats (\EG, headers, inputs) as well as ``functions'' over these objects. For instance, \cref{fig:overview-example} shows an example model defined in the \tool library to test the part of the lookup logic in DNS that determines if a DNS query matches a resource record defined in a zone configuration.


\tool follows a modular approach in defining, synthesizing and composing modules to build a model. Dependencies are specified via the library’s built-in graph API. In the Python code in \cref{fig:overview-example}a, the first line defines a DNS domain name type (\pcode{domain_name}) as a string and limits its size to \pcode{5} characters for testing. It also defines DNS-specific types such as a record type (\pcode{"RecordType"}) as an \pcode{Enum} and a resource record (\pcode{"RR"}) as a \pcode{struct} containing a record type  (\pcode{rtyp}), domain name string (\pcode{name}), and data value (\pcode{rdat}).

The main functionality is split into three modules. The first module \pcode{valid_query} takes a DNS query (\pcode{"query"}) input (simplified for the example by removing the query type) and ensures that its format is valid. It is defined using a \pcode{RegexModule}, which is a pre-defined module type built into \tool. The second module \pcode{ra} (\pcode{"record_applies"}) is the main functionality to test. It is defined as a function \pcode{FuncModule} that takes two arguments, a DNS query (\pcode{query}) and a DNS resource record (\pcode{record}), and returns a single boolean output (\pcode{result}). The function determines if the record is a match for the query. Finally, the third module \pcode{da} (\pcode{"dname_applies"}) is a helper module to assist in the construction of \pcode{ra}. It defines the matching logic specifically for records that have the \pcode{DNAME} record type, which has arguably the most complicated lookup logic among DNS record types.


The user then creates an Eywa \texttt{DependencyGraph} to indicate the interrelations among these modules.
Here we use \pcode{Pipe} to pipe the output of the \pcode{RegexModule} module that produces a valid DNS query to the first input of the \pcode{ra} \pcode{FuncModule} (since it is the first \pcode{Pipe} added to \pcode{ra}). Next, we add a \pcode{CallEdge} to indicate to \tool that the implementation of \pcode{record_applies} can invoke the \pcode{dname_applies} function.

Finally, the last two lines of the figure use the \tool API to synthesize the end-to-end model for the given graph and then generate tests cases, each of which is a list of arguments and the expected result, for example:
{\small
\begin{lstlisting}[style=PythonStyle,xleftmargin=0pt,]
['a.*',{'rtyp':'DNAME','name':'*','rdat':'a.a'},False]
\end{lstlisting}
}

\para{How it works}
\tool first prompts the LLM to implement each function module in C code (circle1) in \cref{fig:overview-example}). It takes into account the input and output argument type definitions and descriptions. Each \texttt{RegexModule} is translated into a separate C function that implements the constraint-checking logic using a pre-defined regex-matching implementation in C. \tool also creates a \pcode{main} function that enables symbolic execution by declaring symbolic inputs for the model that the LLM produces (circle2). 



\subsection{Dealing with imperfect models}
\label{subsec:overview-constraints}

\begin{lrbox}{\codef}
\begin{lstlisting}[style=CStyle]
bool dname_applies(char* query, Record record) {
    int l1 = strlen(query);
    int l2 = strlen(record.name);
    
    // If the DNAME domain name is longer than
    // the domain name, it cannot be a match.
    if (l2 > l1) {
        return false;
    }
    // Compare the domain names in reverse order.
    for (int i = 1; i <= len2; i++) {
        if (query[l1 - i] != record.name[l2 - i]) {
            return false;
        }
    }
    // If the DNAME domain name is equal to the
    // domain name, it is a match.
    if (l2 == l1) {
        return true;
    }
    // If the character before the DNAME
    // domain name is a dot, it is a match.
    if (query[l1 - l2 - 1] == '.') {
        return true;
    }
    return false;
}
\end{lstlisting}
\end{lrbox}

\begin{figure}[t!]
\centering
\begin{tikzpicture}
    \node[inner sep=2pt] (codea) at (0,0) {
      \begin{tikzpicture}
        \node[inner sep=5pt,rounded corners=.1cm,draw=gray,fill=gray!0,thick] (impl) at (0,0) { \usebox\codef };
      \end{tikzpicture}};
    \node[draw, dashed, minimum width=7.1cm, minimum height=2.1cm,draw=red] (highlight) at (0,-1.75) { };
    \node[] (text) at (2.6,-2.6) { Model bug! };
\end{tikzpicture}
\vspace{-2.2em}
\caption{LLM model for the example in \cref{fig:overview-example}.}
\label{fig:overview-example-impl}
\end{figure}


LLMs can make mistakes, and an imperfect model can result in test inputs that are either invalid or otherwise not useful. For our DNS example, the implementation returned by the LLM (slightly simplified for clarity) is shown in \cref{fig:overview-example-impl}. The highlighted region of code for the helper function \pcode{dname_applies} is actually incorrect -- a DNAME record can only apply to a DNS query if it is shorter than the query. 

We address these kinds of issues in two ways. First, we ask the LLM to generate $k$ models rather than just a single one, and we generate tests from all of the models using Klee. In our experiments for DNS described later, for example, we set $k$ to 10. Hence even if some LLM-generated models are incorrect or incomplete, those errors can be compensated for by other generated models. 

Second, as described earlier we employ differential testing to find bugs, comparing the results of multiple implementations on each test case. This means that we do not rely on the LLM-generated model's result, but rather we simply use it to generate inputs that drive the model down different execution paths. As a result, the model itself can also be simpler than would otherwise be required. In our example, we only model the logic that determines whether a resource record applies to a given query, but we need not model the logic of what should be returned upon a successful match.

Further, we have observed that some model mistakes can even be helpful. In the case of the logical error above, this will result in Klee generating an extra test for the case where the lengths of the query and DNAME record domain names are equal, which is a useful corner case. While the LLM-generated model produces the wrong result for this case, the test can still be used to find bugs in real implementations by comparing their results to one another (\EG, see \cref{subsec:case-studies-test-setup}).

\subsection{Finding implementation bugs}
\label{subsec:overview-bugs}


\tool is effective at finding implementation bugs. For example, \tool found a previously unknown bug in the Knot authoritative nameserver implementation~\cite{knot} with the simple LLM-generated model for \dname records shown in \cref{subsec:overview-dns}. For DNS, we craft valid DNS zone files and queries from the test inputs via a postprocessing step that adds necessary resource records (\EG, SOA and NS) and modifies the test's domain names to have the same suffix. Following this approach, \tool created the following zone file from a test input (we add the \texttt{.test.} suffix in this example):
\vspace{0.5em}
\begin{mdframed}[innerleftmargin=0pt,innertopmargin=2pt, innerbottommargin=2pt,tikzsetting={draw=black!90,line width=.6pt},backgroundcolor=white]
    \begin{tabular}{rll}
            $\lDN{test}$   & \soa  &   $\ldots$  \\ 
            $\lDN{test}$ & \ns  &   $\lDN{ns1}{outside}{edu}$ \\ 
            $\lDN{*}{test}$ & \dname & $\lDN{a}{a}{test}$ \\ 
    \end{tabular}
\end{mdframed}
\vspace{0.5em}
The test DNS query \tool generated was $\tuple{\lDN{a}{*}{test}}{\cname}$. 
The \dname record above matches the query, so nameserver implementations are expected to return the \dname record along with a synthesized \cname record, as follows:
\vspace{0.5em}
\begin{mdframed}[innerleftmargin=0pt,innertopmargin=2pt, innerbottommargin=2pt,tikzsetting={draw=black!90,line width=.6pt},backgroundcolor=white]
    \begin{tabular}{rll}
            $\lDN{*}{test}$ & \dname & $\lDN{a}{a}{test}$ \\ 
            $\lDN{a}{*}{test}$ & \cname & $\lDN{a}{a}{a}{test}$ \\ 
    \end{tabular}
\end{mdframed}
\vspace{0.5em}
Knot synthesized the \cname correctly but it generated a new \dname record, as follows:
\vspace{0.5em}
\begin{mdframed}[innerleftmargin=0pt,innertopmargin=2pt, innerbottommargin=2pt,tikzsetting={draw=black!90,line width=.6pt},backgroundcolor=white]
    \begin{tabular}{rll}
            $\lDN{a}{*}{test}$ & \dname & $\lDN{a}{a}{test}$ \\ 
            $\lDN{a}{*}{test}$ & \cname & $\lDN{a}{a}{a}{test}$ \\ 
    \end{tabular}
\end{mdframed}
\vspace{0.5em}
When a DNS resolver receives this response from the Knot DNS nameserver, the resolver will incorrectly determine that the \dname record does not apply to the query and hence produce incorrect behavior.
%
After discovering the bug, we filed the issue on the Knot Gitlab source code, and the developers responded positively and fixed the issue within a week.

\subsection{Limitations of the approach}
\label{subsec:overview-limitations}

For \tool to be effective, the LLM must have a strong understanding of the protocol that it will be modeling. For many protocols (\EG, DNS, BGP, ICMP, \ETC) many LLMs today already understand them well due to the vast amount of knowledge widely available for the protocols. Hence in our running example we only had to provide a single sentence to describe the goal of our \pcode{record_applies} function. For new protocols deployed beyond the LLMs training cutoff date or for proprietary protocols whose specifications are not publicly available, users must provide additional protocol-specific documentation, for instance by fine-tuning the LLM or providing things like protocol RFCs within \tool's prompts.

\begin{figure*}
\centering
\resizebox{!}{.6\columnwidth}{
\begin{tikzpicture}
    \node[style=rectangle, fill=white!5](box1) at (0,0){
        \begin{tikzpicture}
            \node[](user) at(0, 0) {\includegraphics[width=0.7cm]{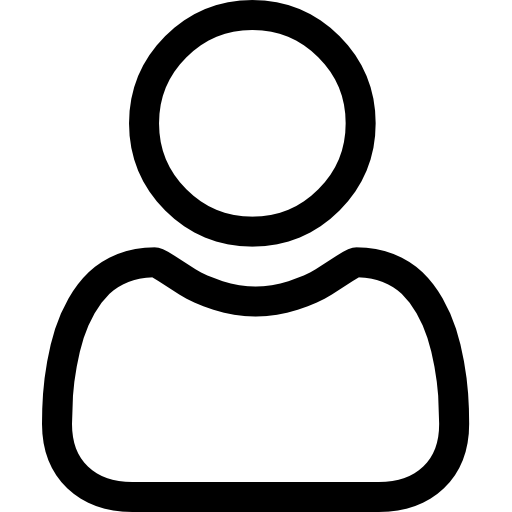}};
            \node[](test_cases) at(0, -0.7) {User};
            \node[fill=blue!10, draw, minimum width=3cm, minimum height=.7cm] (module1) at (3, 2) { Module 1};
            \node[fill=blue!10, draw, minimum width=3cm, minimum height=.7cm] (module2) at (3, 1) { Module 2};
            \node[] (module3) at (3, 0) { ... };
            \node[fill=blue!10, draw, minimum width=3cm, minimum height=.7cm] (modulen) at (3, -0.7) { Module n};
            \node[](graph) at(3, -2.3) {\includegraphics[width=3cm, height=1.8cm]{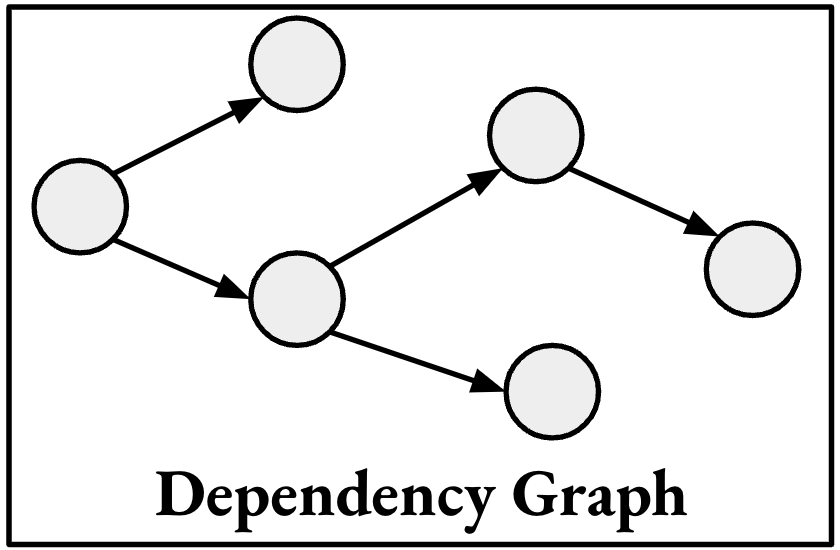}};
            \node[draw, minimum width=2cm, minimum height=.7cm] (eywa) at (6.5, 0) { \tool};
            \node[fill=orange!10, draw, minimum width=2cm, minimum height=.7cm] (prompt1) at (9.5, 2) { Prompt 1};
            \node[fill=orange!10, draw, minimum width=2cm, minimum height=.7cm] (prompt2) at (9.5, 1) { Prompt 2};
            \node[] (prompt3) at (9.5, 0) { ...};
            \node[fill=orange!10, draw, minimum width=2cm, minimum height=.7cm] (promptn) at (9.5, -0.7) { Prompt n};
            \node[fill=magenta!10, draw, minimum width=2cm, minimum height=1cm, align=center] (harness) at (9.5, -2.2) {Symbolic\\Harness};
            \node[](llm1) at(11.7, 2) {\includegraphics[width=0.75cm]{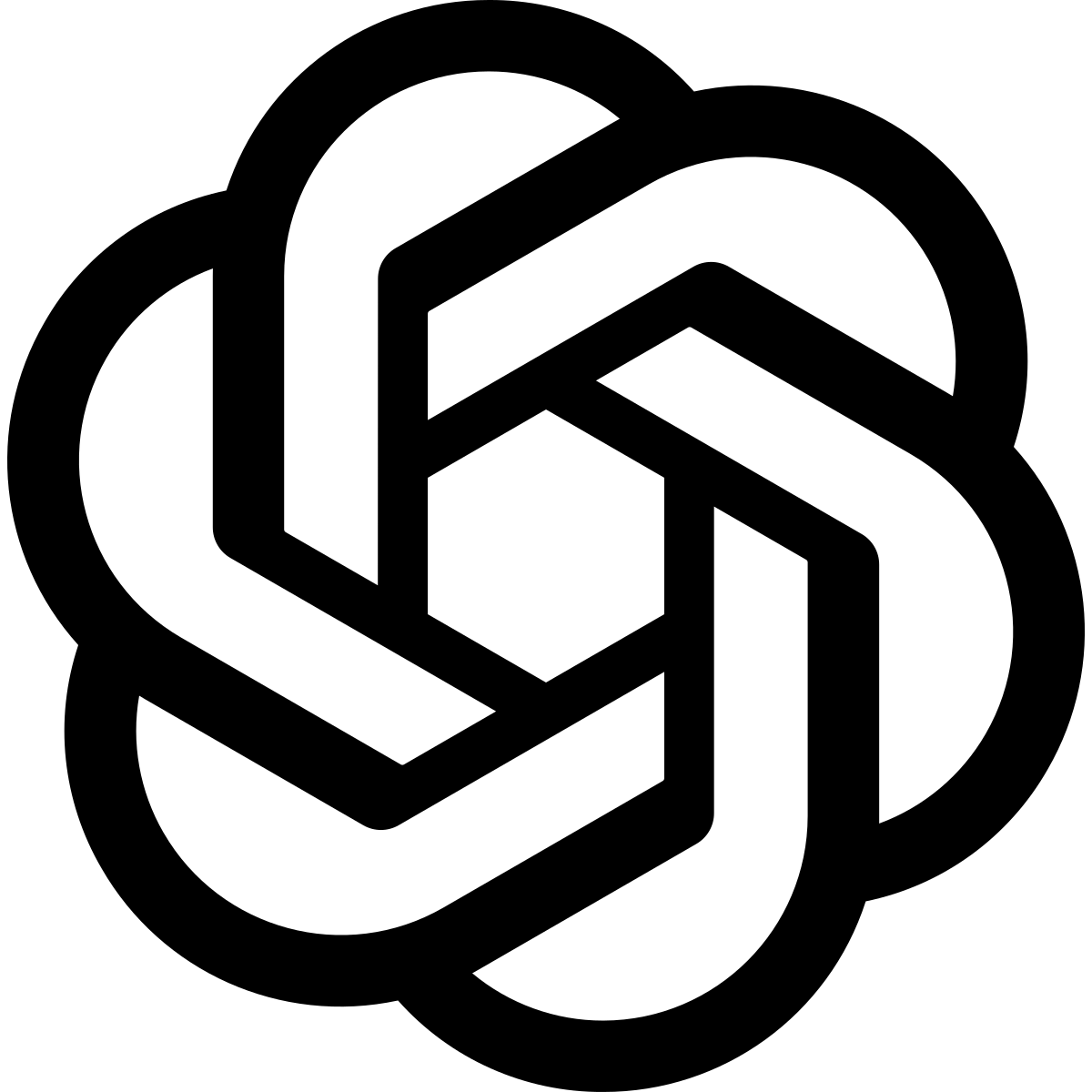}};
            \node[](llm2) at(11.7, 1) {\includegraphics[width=0.75cm]{icons/llm.png}};
            \node[](llmn) at(11.7, -0.7) {\includegraphics[width=0.75cm]{icons/llm.png}};
            \node[fill=pink!10, draw, minimum width=1.5cm, minimum height=.7cm] (code1) at (13.6, 2) { Code 1};
            \node[fill=pink!10, draw, minimum width=1.5cm, minimum height=.7cm] (code2) at (13.6, 1) { Code 2};
            \node[] (code3) at (13.6, 0) { ...};
            \node[fill=pink!10, draw, minimum width=1.5cm, minimum height=.7cm] (coden) at (13.6, -0.7) { Code n};
            \node[fill=yellow!10, draw, minimum width=1.5cm, minimum height=5.5cm, align=center] (code) at (16.2, .2) {Models};
            \node[fill=yellow!10, draw, minimum width=1.5cm, minimum height=5.5cm, align=center] (code) at (16.1, .1) {Models};
            \node[fill=yellow!10, draw, minimum width=1.5cm, minimum height=5.5cm, align=center] (code) at (16, 0) {Models};
            \node[](klee) at(18.2, 0) {\includegraphics[width=1.75cm]{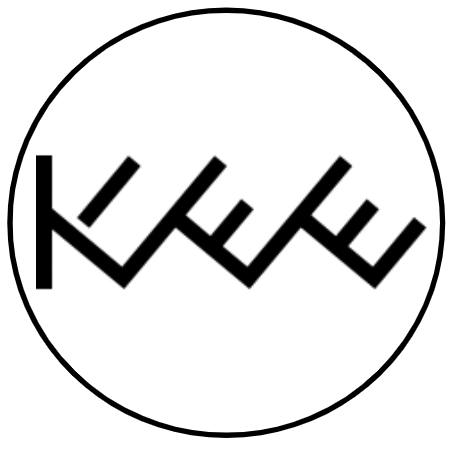}};
            \node[](tests) at(20.5, 0) {\includegraphics[width=1.2cm]{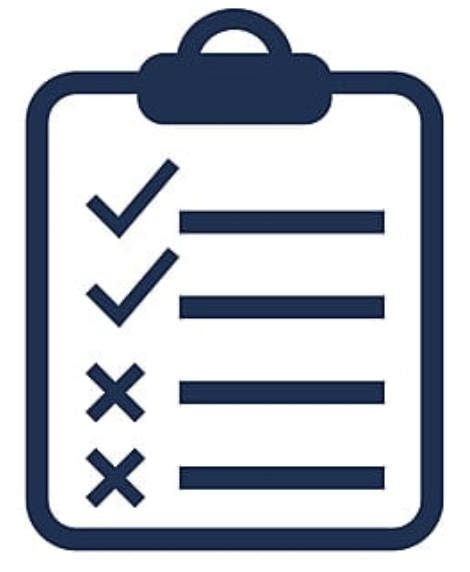}};
            \node[](test_cases) at(20.5, -1) {Test cases};
            \node[draw, dashed, very thick, minimum width=10cm, minimum height=5.8cm] (boundingbox) at (9.8,-0.3) {};
            \node[](test_cases) at(13.6, -2.9) {\textbf{Run \emph{k} times}};
            
            \draw[->] (user.east) -- (module1.west);
            \draw[->] (user.east) -- (module2.west);
            \draw[->] (user.east) -- (modulen.west);
            \draw[->] (user.east) -- (1.5,-2.3);
            \draw[->] (module1.east) -- (eywa.west);
            \draw[->] (module2.east) -- (eywa.west);
            \draw[->] (modulen.east) -- (eywa.west);
            \draw[->] (4.5,-2.3) -- (eywa.west);
            \draw[->] (eywa.east) -- (prompt1.west);
            \draw[->] (eywa.east) -- (prompt2.west);
            \draw[->] (eywa.east) -- (promptn.west);
            \draw[->] (eywa.east) -- (harness.west);
            \draw[->] (prompt1.east) -- (llm1.west);
            \draw[->] (prompt2.east) -- (llm2.west);
            \draw[->] (promptn.east) -- (llmn.west);
            \draw[->] (llm1.east) -- (code1.west);
            \draw[->] (llm2.east) -- (code2.west);
            \draw[->] (llmn.east) -- (coden.west);
            \draw[->] (code1.east) -- (15.2,2);
            \draw[->] (code2.east) -- (15.2,1);
            \draw[->] (coden.east) -- (15.2,-0.7);
            \draw[->] (harness.east) -- (15.2,-2.2);
            \draw[->] (code.east) -- (klee.west);
            \draw[->] (klee.east) -- (tests.west);
            
        \end{tikzpicture}
    };
\end{tikzpicture}}
\vspace{-1.3em}
\caption{Architecture of \tool. It takes modules and dependencies and produces \emph{k} different protocol models.}
\label{fig:system-architecture}
\end{figure*}

%
%
%
%

\section{The \tool Library and Runtime}
\label{sec:system}

In this section we describe \tool's design and runtime.

\subsection{\tool Architecture}
\label{subsec:system-architecture}

\tool's architecture comprises several interacting components whose worksflow is illustrated in \cref{fig:system-architecture}. To start, the user (extreme left) uses the \tool library to define protocol modules (functions) with arguments and return values and composes them together through a dependency graph, as shown in the example from \cref{fig:overview-example}. From these inputs, \tool then internally invokes its \emph{Prompt Generator} and \emph{Symbolic Compiler} to produce the model.

The \textit{Prompt Generator} builds an LLM prompt for each \texttt{FuncModule} declared by the user. It invokes the LLM on each prompt to produce a corresponding C function. 
%
The \textit{Symbolic Compiler} uses the dependency graph to build the  \textit{Symbolic Harness}, which is the \texttt{main} function that represents the full model.  The function's inputs are declared symbolic using the Klee API, to make the code amenable to symbolic execution.  The \textit{Symbolic Harness} also produces the C functions for pre-defined modules like the \texttt{RegexModule}. 
The complete model consists of the outputs of both the \textit{Prompt Generator} and \textit{Symbolic Compiler}. This code synthesis process is repeated $k$ times to generate $k$ different models. Finally, the \textit{Test Generator} invokes Klee on these models, extracts the generated test cases from Klee, and translates them back into Python data structures.

\subsection{\tool Library}
\label{subsec:system-library-design}


The Eywa library facilitates building models as typed functions that accept a set of arguments and return a result. This design is general and is applicable to many kinds of protocols. For instance, to model the logic of a stateful network protocol like SMTP, one could create a function that takes (1) an SMTP message representing a received email or command, and (2) the current state of the SMTP protocol for the host handling the message, and returns (3) the appropriate SMTP response or action to take and (4) the updated state.


\begin{figure}[t!]
\centering
\setlength\tabcolsep{3pt} 
\begin{tabular}{ll}
    \toprule
    \textbf{Example Feature} & \textbf{Description} \\ 
    \midrule
    \pcode{eywa.Bool()} & Boolean value \\
    \pcode{eywa.Char()} & Character value \\
    \pcode{eywa.String(maxsize=5)} & String of size \\
    \pcode{eywa.Int(bits=5)} & 5-bit unsigned int \\
    \pcode{eywa.Enum("name",["A", "NS"])} & Enum value \\
    \pcode{eywa.Array(Bool(), 3)} & Array of 3 bools \\
    \pcode{eywa.Struct("name",dst=Int())} & Struct with 1 field \\
    \pcode{eywa.Alias("result",Bool())} & Type alias \\
    \pcode{eywa.Arg("name",Int(),"desc")} & Function arg \\
    \pcode{eywa.FuncModule("name","desc",args)} & Function \\
    \bottomrule
\end{tabular}
\vspace{-1em}
\caption{Summary of \tool types and abstractions.}
\label{fig:system-types-summary}
\vspace{-\baselineskip}
\end{figure}

A summary of the main modeling abstractions are shown in \cref{fig:system-types-summary}. The library allows for the creation of function arguments with standard types such as booleans, characters, strings, fixed bit-width integers, enums, arrays, structs, as well as type aliases that allow for associating custom names with types (\EG, to help the LLM understand a type's meaning). For types with potentially unbounded size (\EG, \pcode{eywa.String()}), users must also provide hints to bound the size of that type to limit the size and number of test cases that \tool will produce.
Users create functions by giving them a name, a natural language description of their purpose, and a list of arguments. Each argument also has a name and description as well as an associated type.


\subsection{\tool Support for Modularity}
\label{subsec:system-modularity}
As discussed in Section \cref{subsec:overview-dns}, there are two kinds of modules available in \tool. A \pcode{FuncModule} is initialized by providing a natural language description of its content. The code is then written by an LLM through a prompt generated by \tool. \tool also includes a built-in \pcode{RegexModule} since regex filtering is a common need for protocols, but users can provide their own modules as well for specialized functionality for which they want full control.

Dependency graphs determine how modules are connected to form a full model. \tool's dependency graphs support two kinds of edges. A \pcode{Pipe} is used for \emph{sequential composition}, where the output of one module is the input of another. A \pcode{CallEdge} indicates that the implementation of one module may invoke another one and thereby supports a form of \emph{decomposition}: a complex module can be created through multiple LLM prompts. If a module $m_1$ is declared to call another module $m_2$, then the C function prototype for $m_2$ is provided when the LLM is asked to synthesize $m_1$.  A separate LLM invocation will produce the implementation of $m_2$. See \cref{sec:appendix-regex,appendix:graph-api} for further discussion.


\subsection{\tool Prompt Generator}
\label{subsec:system-prompt-generator}

Given a set of modules along with the dependency graph for a model, \tool invokes its \textit{Prompt Generator} to build an LLM prompt that will lead the LLM to implement each module and then proceeds to combine them into one model. For each module, \tool builds two prompts: a user prompt that frames the implementation task as a completion problem, and a system prompt to guide the behavior of the LLM.

Consider again the example model defined using the \tool library in \cref{fig:overview-example}. 

\begin{lrbox}{\codeg}
\begin{lstlisting}[style=CStyle]
#include <stdint.h>
...
typedef enum {
    A, AAAA, NS, TXT, CNAME, DNAME, SOA
} RecordType;

typedef struct {
    RecordType record_type;
    String name;
    String rdata;
} Record;

// If a DNS record matches a query.
// ...
bool dname_applies(char* query, Record record);

// If a DNS record matches a query.
// Parameters:
//     query: A DNS query domain name.
//     dname: A DNS record.
// Return Value:
//     If the DNS record matches the query.
bool record_applies(char* query, Record record) {
\end{lstlisting}
\end{lrbox}

\begin{figure}[t!]
\centering
\begin{tikzpicture}
    \node[inner sep=2pt] (codea) at (0,0) {
      \begin{tikzpicture}
        \node[inner sep=5pt,rounded corners=.1cm,draw=gray,fill=gray!0,thick] (impl) at (0,0) { \usebox\codeg };
      \end{tikzpicture}};
\end{tikzpicture}
\vspace{-1em}
\caption{LLM prompt for the example in \cref{fig:overview-example}.}
\label{fig:overview-prompt}
\vspace{-\baselineskip}
\end{figure}

\para{User Prompt} To create the user prompt, \tool translates each of the user-defined types into C data types, creates the C function signatures using these types, and adds a function documentation string from the user descriptions. We show the resulting prompt that is generated for \pcode{record_applies} of \cref{fig:overview-example} in \cref{fig:overview-prompt}. From this prompt, the LLM will naturally predict (complete) the rest of the function. 

\para{System Prompt} \tool also generates a system prompt (for GPT4). It provides additional guidance to help ensure the LLM generates valid code, including the following:
\begin{itemize}
    \item Describing the task to implement the C function provided by the user prompt.
    \item Requiring that the LLM only add import statements and not remove existing imports.
    \item Requiring that the LLM not delete or modify any of the user-defined type definitions.
    \item Requiring that the LLM not add its own \ccode{main()} function and instead just implement the function provided.
    \item Requiring that the LLM not use certain C functions that are not amenable to symbolic execution.
    \item Giving an example of a valid input and output.
\end{itemize}
%
To see additional examples, refer to \cref{appendix:graph-api,appendix:system-prompt}.

\subsection{\tool Symbolic Compiler}
\label{subsec:system-symbolic-compiler}

The \textit{Symbolic Compiler} generates the \textit{Symbolic Harness}, which is the \ccode{main} function used for symbolic execution.  We use the \ccode{klee_make_symbolic} command to build symbolic inputs for every base type (\ccode{bool}, \ccode{char}, \ccode{int}, \ccode{enum} \ETC) and then construct values of more complex types from these base types.  For example, a struct will be initialized with symbolic values for each of its fields.



\subsection{\tool Test Generator}
\label{subsec:system-test-generator}
The final component to \tool is the \textit{Test Generator}, which is responsible for running Klee and translating the results back into Python values for the user. The \textit{Test Generator} assembles the outputs from the \textit{Prompt Generator} and \textit{Symbolic Compiler} into a single program and then runs the \texttt{clang} compiler to build LLVM bytecode. It then executes Klee on the result with the user-provided timeout (if any). For isolation, both of these tasks are executed in a separate docker container and any errors, including compile errors, are reported to the user as feedback. Users can use this feedback to update their model to assist the LLM or provide additional context.

The result of running Klee is a set of test inputs that assign each base symbolic C variable to a value. The \textit{Test Generator} then serializes these values back to \tool, which walks over the base values and reconstructs any richer types (\EG, struct, array) from these values. 
\tool also captures the result value of each test case and includes that information.

%
%
%
%

\section{Evaluation}
\label{sec:evaluation}

\subsection{Implementation}
\label{sec:implementation}

The \tool library is implemented in around 2K lines of Python code with an additional 200 lines of C code. For its language model, it currently uses GPT 4 hosted on the Azure OpenAI service~\cite{azure-openai}. When generating models, \tool allows the user to specify the LLM temperature value $\tau$ between 0.0 and 1.0 as well as the number of C implementations $k$ to generate for each model. To improve test coverage, we generate $k$ implementations and then aggregate the resulting test cases. For our experiments, we use a $k = 10$ and temperature $\tau = 0.6$ (see \cref{sec:appendix-hyperparameter}). After assembling each model from the LLM and symbolic test harness, we attempt to compile it and invoke Klee in a Docker container, and skip the implementation in the event of a compilation error.



\subsection{Methodology and Setup}

Alongside DNS, we also experimented with BGP and SMTP, each with multiple implementations that we use for differential testing, summarized in \cref{tab:eval-implementations}. For each protocol, we build several models with \tool in order to test various aspects of the protocol. \cref{tab:eval-tests-and-results} presents all the models we created with \tool for the purpose of testing. The table shows the number of lines of Python code that were needed to define each model in \tool (similar to \cref{fig:overview-example}). The table also gives the range (both the minimum and maximum) of lines of code in C that \tool generates across the $k$ model implementations. Finally, we show the total number of unique test cases for each model returned by \tool after running Klee on the generated C code. This includes the union of all unique test cases across the $k$ different implementations. 

\begin{table}[t!]
    \centering
    \begin{tabular}{ll}
        \toprule
        \textbf{Protocol} & \textbf{Tested Implementations} \\
        \toprule
        \multirow{3}{*}{\textsc{DNS}}
            & \textsc{bind}~\cite{bind9}, \textsc{coredns}~\cite{coredns}, \textsc{gdnsd}~\cite{gdnsd}, \textsc{nsd}~\cite{nsd} \\ 
            & \textsc{hickory}~\cite{trustdns}, \textsc{knot}~\cite{knot}, \textsc{powerdns}~\cite{PowerDNS} \\
            & \textsc{technitium}~\cite{technitium}, \textsc{yadifa}~\cite{yadifa}, \textsc{twisted}~\cite{twisted} \\
        \addlinespace[.5ex]
        \hdashline[1pt/1pt]
        \addlinespace[.5ex]
        \textsc{BGP} & \textsc{frr}~\cite{frr}, \textsc{gobgp}~\cite{gobgp}, \textsc{batfish}~\cite{batfish}  \\
        \addlinespace[.5ex]
        \hdashline[1pt/1pt]
        \addlinespace[.5ex]
        \textsc{SMTP} & \textsc{aiosmtpd}~\cite{aiosmtpd}, \textsc{smtpd}~\cite{smtpd}, \textsc{opensmtpd}~\cite{opensmtpd} \\
        \bottomrule
    \end{tabular}
    \caption{Protocol implementations tested by \tool.}
    \label{tab:eval-implementations}
    \vspace{-2em}
\end{table}

\begin{table}[t!]
    \centering
    \begin{tabular}{llccc}
        \toprule
        \textbf{Protocol} & \textbf{Model} & \textbf{LOC} & \textbf{LOC} & \textbf{Tests}\\
        \textbf{} & \textbf{} & \textbf{(Python)} & \textbf{(C)} & \textbf{}\\
        \toprule
        \textsc{DNS} & \textsc{cname} & 21 & 222 / 246 & 435\\
        \textsc{DNS} & \textsc{dname} & 23 & 209 / 230 & 269 \\
        \textsc{DNS} & \textsc{wildcard} & 23 & 210 / 238 & 470 \\
        \textsc{DNS} & \textsc{ipv4} & 21 & 209 / 229 & 515 \\
        \textsc{DNS} & \textsc{fulllookup} & 26 & 487 / 510 & 12,281 \\
        \textsc{DNS} & \textsc{rcode} & 26 & 487 / 510 & 26,617 \\
        \textsc{DNS} & \textsc{auth} & 26 & 477 / 504 & 31,411 \\
        \textsc{DNS} & \textsc{loop} & 26 & 474 / 489 & 31,453 \\
        \addlinespace[.5ex]
        \hdashline[1pt/1pt]
        \addlinespace[.5ex]
        \textsc{BGP} & \textsc{confed} & 22 & 189 / 202 & 957 \\

        \textsc{BGP} & \textsc{rr} & 16 & 59 / 76 & 36 \\
        \textsc{BGP} & \textsc{rmap-pl} & 48 & 150 / 162 & 400\\
        \textsc{BGP} & \textsc{rr-rmap} & 48 & 341 / 366 & 7147 \\
        \addlinespace[.5ex]
        \hdashline[1pt/1pt]
        \addlinespace[.5ex]
        \textsc{SMTP} & \textsc{server} & 26 & 245 / 252 & 80\\
        \bottomrule
    \end{tabular}
    \caption{Models, lines of code (min / max) , number of tests generated for DNS, BGP and SMTP with \tool.}
    \label{tab:eval-tests-and-results}
    \vspace{-2em}
\end{table}


\para{Overview of models} We constructed eight models of the DNS using \tool (shown in \cref{tab:eval-tests-and-results}) and tested different parts of the protocol for authoritative nameservers. To better understand when \tool performs well and when it does not, we attempted to craft a diverse set of models. We created (1) simple models that test matching behavior for individual DNS records, (2) end-to-end models that capture the entire DNS lookup behavior, and (3) specialized models to target unusual or corner case behavior.

Specifically, the first four models (\textsc{cname}, \textsc{dname}, \textsc{wildcard}, \textsc{ipv4}) each determines whether a DNS query matches a single record type in a DNS zone. 
%
The \textsc{fulllookup} model implemements the entire DNS authoritative lookup procedure for a query and a zone file. 
Two other models, \textsc{authoritative} and \textsc{rcode}, are similar to \textsc{fulllookup} but return only a part of the DNS response -- in particular, the authoritative flag and return code respectively, rather than the full DNS response.
Finally, we include a \textsc{loop} model that tests a corner case of the DNS protocol. In particular, we ask \tool to create a model that counts the number of times a DNS query is rewritten for a given zone file. This forces \tool to explore inputs that result in recursively looking up an answer multiple times or even indefinitely.
These models have a greater number of test cases than \textsc{fulllookup} because of the same Klee timeout for all of them. However, \textsc{fulllookup} is more complex and requires more time to generate a test case.

For the Border Gateway Protocol (BGP), we modeled several features and their combinations.
Specifically our models capture the logic of BGP confederations (\textsc{confed}), how route-reflector servers interact with clients and non-clients and external ASes (\textsc{rr}), how route-maps and prefix lists process BGP route advertisements (\textsc{rmap-pl}), and the combination of route reflectors with route-maps (\textsc{rr-rmap}). 

Finally, for the Simple Mail Transfer Protocol (SMTP), we created a  model of an SMTP server. The model responds to incoming messages/requests depending on the server state, including emitting an appropriate error message for invalid requests. The model is implemented as a function that takes two parameters, state and input, and returns the corresponding server response. \cref{fig:smtp-prompt} shows the prompt generated by \tool and \cref{fig:smtp-code} shows the resulting code from the LLM. 

\begin{lrbox}{\codeg}
\begin{lstlisting}[style=CStyle]
#include <stdint.h>
#include <stdbool.h>
...
typedef enum { INITIAL, HELO_SENT, 
               EHLO_SENT, MAIL_FROM_RECEIVED, 
               RCPT_TO_RECEIVED, DATA_RECEIVED, 
               QUITTED } State;
// A function that takes the current state of  
// the SMTP server, the input string, updates 
// the state and returns the output response. 
//
// Parameters:
//     state: Current state of the SMTP server
//     input: Input string
// Return Value:
//     Output string
char* smtp_server_resp(State state, char* input) {
\end{lstlisting}
\end{lrbox}

\begin{figure}[t!]
\centering
\begin{tikzpicture}
    \node[inner sep=2pt] (codea) at (0,0) {
      \begin{tikzpicture}
        \node[inner sep=5pt,rounded corners=.1cm,draw=gray,fill=gray!0,thick] (impl) at (0,0) { \usebox\codeg };
      \end{tikzpicture}};
\end{tikzpicture}
\vspace{-2em}
\caption{First prompt for the SMTP server Model.}
\label{fig:smtp-prompt}
\vspace{-\baselineskip}
\end{figure}

\begin{lrbox}{\codeg}
\begin{lstlisting}[style=CStyle]
char* smtp_server_response(State state, char* input) {
    char* response = malloc(512 * sizeof(char));
    if(state == INITIAL) {
        if(strcmp(input, "HELO") == 0) {
            strcpy(response, "250 Hello");
            state = HELO_SENT;
        } else if(strcmp(input, "EHLO") == 0) {
            strcpy(response, 
            "250-Hello\n250-SIZE 512\n250-8BITMIME\n250 OK");
            state = EHLO_SENT;
        } else {
            strcpy(response, "503 Bad sequence of commands");
        }
    } else if(state == HELO_SENT || state == EHLO_SENT) {
        if(strncmp(input, "MAIL FROM:", 10) == 0) {
            strcpy(response, "250 OK");
            state = MAIL_FROM_RECEIVED;
        } else {
            strcpy(response, "503 Bad sequence of commands");
        }
    } else if(state == MAIL_FROM_RECEIVED) {
        if(strncmp(input, "RCPT TO:", 8) == 0) {
            strcpy(response, "250 OK");
            state = RCPT_TO_RECEIVED;
        } else {
            strcpy(response, "503 Bad sequence of commands");
        }
    } else if(state == RCPT_TO_RECEIVED) {
        if(strcmp(input, "DATA") == 0) {
            strcpy(response, "354 End with <CR><LF>.<CR><LF>");
            state = DATA_RECEIVED;
        } else {
            strcpy(response, "503 Bad sequence of commands");
        }
    } else if(state == DATA_RECEIVED) {
        if(strcmp(input, ".") == 0) {
            strcpy(response, "250 OK");
            state = INITIAL; 
        } else {
            strcpy(response, "");
        }
    } else if(state == QUITTED) {
        strcpy(response, "221 Bye");
        state = INITIAL;
    } else {
        strcpy(response, "500 error, command unrecognized");
    }
    return response;
}
\end{lstlisting}
\end{lrbox}

\begin{figure}[t!]
\centering
\resizebox{\columnwidth}{!}{%
\begin{tikzpicture}
    \node[inner sep=2pt] (codea) at (0,0) {
      \begin{tikzpicture}
        \node[inner sep=5pt,rounded corners=.1cm,draw=gray,fill=gray!0,thick] (impl) at (0,0) { \usebox\codeg };
      \end{tikzpicture}};
\end{tikzpicture}
}
\vspace{-2em}
\caption{\tool generated code for SMTP server Model.}
\label{fig:smtp-code}
\vspace{-\baselineskip}
\end{figure}



\para{Overview of testing setup}
Seven of the ten DNS implementations that we tested were also tested by SCALE~\cite{scale}. For those seven we additionally tested both the old versions (prior to bug fixes) of each implementation along with current versions. Doing so allowed us to identify the overlap between bugs found by \tool and SCALE as well as any new bugs that were yet undiscovered.
%
To test each DNS implementation we use differential testing as described in \cref{fig:overview-example}. We created scripts to initialize a Docker~\cite{Docker} container for each DNS implementation and version. Each container serves a single zone file as an authoritative zone, and we used the \texttt{dnspython}~\cite{dnspython} library to construct DNS queries and send them to each container. After receiving the response, we compared the answers to find which parts differ (\EG, answer, authoritative section, flags, additional section, return code).

Since many tests can trigger the same bug, to make it easier to identify unique root causes, we take each implementation whose response is not part of the majority group and classify the ``reason'' for the disagreement as a tuple that abstracts the relevant parts that differ. For instance, a response that returns code \texttt{NXDOMAIN} instead of \texttt{NOERROR} might have the tuple (\textsc{coredns}, \texttt{rcode}, \texttt{NXDOMAIN}, \texttt{NOERROR}). For each such unique tuple we manually inspected the failing tests and reported any bugs found with the implementation maintainers.

For the BGP implementations we set up a Docker network containing three nodes R1, R2 and R3 connected in series. We run the BGP implementation under test on R2 and R3 but run an ExaBGP~\cite{exabgp} instance on R1 so that we can inject routes into R2. We then check the BGP routing tables on both R2 and R3. For Batfish, which is a BGP simulator rather than an implementation, we set up a simulated network of three nodes in similar fashion.
We wrote test translators for all three implementations, and then performed differential testing on the three implementations. Error triaging is done in the same way as for the DNS experiments.

Finally for SMTP, as described earlier each test case is an (state, input) pair. Hence running each test case requires us to first drive each SMTP implementation into the desired state. To achieve this, \tool uses \emph{another LLM call} to transform the generated SMTP server code into a state transition graph (see \cref{fig:state-graph-prompt-response}). For each test case, \tool  uses breadth-first search (BFS) on the state graph to determine an input sequence that will transition the server from the initial state to the desired state for that test, and this input sequence is prepended to the test's input. The server implementations are hosted at IP 127.0.0.1 and port 8025; another process on the same machine generates the input requests for the servers. After executing each test case, the server is reset to its initial state.

\begin{lrbox}{\codeg}
\begin{lstlisting}[style=CStyle,
    keywords={},
    morekeywords={User, Prompt, GPT, Response},
    keywordstyle=\color{blue}\bfseries, 
    stringstyle=\color{red}, 
    commentstyle=\color{gray}\itshape
]
User Prompt:
Create a python dictionary that maps the state 
transitions: (state,input) --> state 
as per the following C code snippet:

smtp_server_response(State state, char* input) {
    <SMTP server code generated by LLM>
}

Output format: 
A python dictionary like 
{(state1, input1): state2, 
(state3, input2): state4, ...}
    
GPT Response:
Here is the Python dictionary that maps the 
state transitions:

```python
state_transitions = {
    ("INITIAL", "HELO"): "HELO_SENT",
    ("INITIAL", "EHLO"): "EHLO_SENT",
    ("HELO_SENT", "MAIL FROM:"): "MAIL_FROM_RECEIVED",
    ("EHLO_SENT", "MAIL FROM:"): "MAIL_FROM_RECEIVED",
    ("MAIL_FROM_RECEIVED", "RCPT TO:"): 
    "RCPT_TO_RECEIVED",
    ("RCPT_TO_RECEIVED", "DATA"): "DATA_RECEIVED",
    ("HELO_SENT", "QUIT"): "QUITTED",
    ("EHLO_SENT", "QUIT"): "QUITTED",
    ("MAIL_FROM_RECEIVED", "QUIT"): "QUITTED",
    ("RCPT_TO_RECEIVED", "QUIT"): "QUITTED",
    ("DATA_RECEIVED", "QUIT"): "QUITTED",
}
```
\end{lstlisting}
\end{lrbox}

\begin{figure}[t!]
\centering
\begin{tikzpicture}
    \node[inner sep=2pt] (codea) at (0,0) {
      \begin{tikzpicture}
        \node[inner sep=5pt,rounded corners=.1cm,draw=gray,fill=gray!0,thick] (impl) at (0,0) { \usebox\codeg };
      \end{tikzpicture}};
\end{tikzpicture}
\vspace{-2em}
\caption{Second LLM prompt and response to generate the SMTP state graph as a Python dictionary.}
\label{fig:state-graph-prompt-response}
\vspace{-\baselineskip}
\end{figure}

\begin{table*}[t!]
    \setlength{\tabcolsep}{3pt}
    \centering
    \begin{tabular}{llllccc}
        \toprule
        \textbf{Protocol} & \textbf{Implementation} & \textbf{Description} & \textbf{Bug Type} & \textbf{New bug?}
        \\
        \midrule
        DNS & \textsc{bind}       & Sibling glue record not returned.                     & Wrong Additional  & \fullmoon \\
        DNS & \textsc{bind}       & Inconsistent loop unrolling.                          & Wrong Answer      & \newmoon \\
        DNS & \textsc{coredns}    & Wildcard CNAME and DNAME loop.                        & Server Crash      & \fullmoon \\
        DNS & \textsc{coredns}    & Sibling glue record not returned.                     & Wrong Additional  & \fullmoon \\
        DNS & \textsc{coredns}    & Returns SERVFAIL yet gives an answer.                 & Wrong Answer      & \newmoon \\
        DNS & \textsc{coredns}    & Missing record for CNAME loop.                        & Wrong Answer      & \newmoon \\
        DNS & \textsc{coredns}    & Returns a non-existent out-of-zone record.            & Wrong Answer      & \newmoon \\
        DNS & \textsc{coredns}    & Wrong RCODE when `*' is in RDATA.                     & Wrong Return Code & \fullmoon \\
        DNS & \textsc{coredns}    & Wrong RCODE for empty non-terminal wildcard.          & Wrong Return Code & \newmoon \\
        DNS & \textsc{gdnsd}       & Sibling glue record not returned.                    & Wrong Additional  & \fullmoon \\
        DNS & \textsc{hickory}    & Wildcard CNAME and DNAME loop.                        & Server Crash      & \fullmoon \\
        DNS & \textsc{hickory}    & Incorrect handling of out-of-zone record.             & Wrong Answer      & \newmoon \\
        DNS & \textsc{hickory}    & Wildcard match only one label.                        & Wrong Answer      & \fullmoon \\
        DNS & \textsc{hickory}    & Wrong RCODE for empty non-terminal wildcard.          & Wrong Return Code & \newmoon \\
        DNS & \textsc{hickory}    & Wrong RCODE when `*' is in RDATA.                     & Wrong Return Code & \newmoon \\
        DNS & \textsc{hickory}    & Glue records returned with authoritative flag.        & Wrong Flags       & \fullmoon \\
        DNS & \textsc{hickory}    & Authoritative flag set for zone cut NS records.       & Wrong Flags       & \fullmoon \\
        DNS & \textsc{knot}       & DNAME record name replaced by query.                  & Wrong Answer      & \newmoon \\
        DNS & \textsc{knot}       & Wildcard DNAME leads to wrong answer.                 & Wrong Answer      & \newmoon \\
        DNS & \textsc{knot}       & Error in DNAME-DNAME loop Knot test.                  & Faulty Knot Test  & \fullmoon \\
        DNS & \textsc{knot}        & DNAME not applied recursively.                       & Wrong Answer      & \fullmoon \\
        DNS & \textsc{knot}        & Incorrect record synthesis when `*' is in query.     & Wrong Answer      & \fullmoon \\
        DNS & \textsc{nsd}        & DNAME not applied recursively.                        & Wrong Answer      & \fullmoon \\
        DNS & \textsc{nsd}        & Wrong RCODE when `*' is in RDATA.                     & Wrong Return Code & \fullmoon \\
        DNS & \textsc{powerdns}   & Sibling glue record not returned due to wildcard.     & Wrong Additional  & \newmoon \\
        DNS & \textsc{technitium} & Sibling glue record not returned.                     & Wrong Additional  & \fullmoon \\
        DNS & \textsc{technitium} & Synthesized wildcard instead of applying DNAME.       & Wrong Answer      & \newmoon \\
        DNS & \textsc{technitium} & Invalid wildcard match.                               & Wrong Answer      & \fullmoon \\
        DNS & \textsc{technitium} & Nested wildcards not handled correctly.               & Wrong Answer      & \newmoon \\
        DNS & \textsc{technitium} & Duplicate records in answer section.                  & Wrong Answer      & \fullmoon \\
        DNS & \textsc{technitium} & Wrong RCODE for empty nonterminal wildcard.           & Wrong Return Code & \newmoon \\
        DNS & \textsc{twisted}    & Empty answer section with wildcard records.           & Wrong Answer      & \fullmoon \\
        DNS & \textsc{twisted}    & Missing authority flag and empty authority section.   & Wrong Flags       & \fullmoon \\
        DNS & \textsc{twisted}    & Wrong RCODE for empty nonterminal wildcard.         & Wrong Return Code & \newmoon \\
        DNS & \textsc{twisted}    & Wrong RCODE when `*' is in RDATA.                     & Wrong Return Code & \fullmoon \\
        DNS & \textsc{yadifa}     & CNAME chains are not followed.                        & Wrong Answer      & \fullmoon \\
        DNS & \textsc{yadifa}     & Missing record for CNAME loop.                        & Wrong Answer      & \newmoon \\
        DNS & \textsc{yadifa}     & Wrong RCODE for CNAME target.                         & Wrong Return Code & \fullmoon \\
        \addlinespace[.5ex]
        \hdashline[1pt/1pt]
        \addlinespace[.5ex]
        BGP & \textsc{FRR}        & Prefix list matches mask greater than or equals.      & Wrong Policy      & \fullmoon \\
        BGP & \textsc{FRR}        & Confederation sub AS equal to peer AS.                & Wrong Policy     & \newmoon \\
        BGP & \textsc{FRR}        & Replace-AS not working with confederations.           & Wrong Policy      & \newmoon \\
        BGP & \textsc{GoBGP}      & Prefix set match with zero masklength but nonzero range. & Wrong Policy   & \fullmoon \\
        BGP & \textsc{GoBGP}      & Confederation sub AS equal to peer AS.                & Wrong Policy      & \newmoon \\
        BGP & \textsc{Batfish}    & Local preference not reset for EBGP neighbor.         & Wrong Policy      & \newmoon \\
        BGP & \textsc{Batfish}    & Confederation sub AS same as peer AS.                 & Wrong Policy      & \newmoon\\
        \addlinespace[.5ex]
        \hdashline[1pt/1pt]
        \addlinespace[.5ex]
        SMTP & \textsc{AioSMTPD}  & Server accepting request without appropriate headers  & Input Validation  & \newmoon \\
        \bottomrule
    \end{tabular}
    \vspace{.6em}
    \caption{The bugs found in the DNS, BGP, and SMTP protocol implementations tested by \tool.}
    \label{tab:eval-bugs}
    \vspace{-\baselineskip}
\end{table*}

\subsection{Results}
\label{subsec:case-studies-test-setup}

In this section, we describe the results of our experiments with \tool on DNS, BGP, and SMTP implementations, including quantitative and qualitative evidence of its practical utility and benefit over prior approaches. 

\para{1. \tool produces tests quickly} The running time for \tool is dominated by the time to test the implementations. Each LLM query took under 20 seconds to complete and could be reduced by using faster LLMs. For test generation using Klee, the time taken varies based on the complexity of the case. For the initial four models of DNS and for the SMTP server model Klee completes the process in approximately 5-10 seconds. However, for the other DNS models which are more complex, Klee consistently hits the 5-minute timeout that we use. On the other hand, all our BGP models are bounded in size so Klee always terminated within 5-10 seconds (well before the timeout). 

\para{2. \tool produces high-quality models} We manually inspected \tool's generated implementations to assess their quality. For each of the models from \cref{tab:eval-tests-and-results}, we found that \tool produced models that largely capture the intended protocol semantics faithfully. For instance, for the \textsc{DNS fulllookup} task, which is the most challenging DNS task since it asks for a full model of DNS nameserver lookup, the LLM correctly implemented the behavior of each record type including DNAME (suffix rewrite), CNAME (exact rewrite), and wildcard (partial match). The LLM did not perfectly implement the ``closest encloser'' semantics of DNS lookup (doing so would require creating a complex data structure), instead typically performing a ``first-match'' semantics by iterating through the list of zone records sequentially. While technically incorrect, this 
is a close approximation of the true behavior and produced an effective set of test cases when combined with symbolic execution. 

For BGP, \tool's models again were largely accurate. 
With careful prompt engineering, we could even force the LLM to cover missed corner cases. However, for BGP confederations, we observed that the LLM struggled to update the AS paths properly, despite trying multiple prompt variations. However, because we employ differential testing and ignore model results, the model's logic still produced useful test cases.  Finally, for the SMTP server, \tool almost always generated a perfect model.

In several cases, the LLM originally implemented a flawed model due to its misunderstanding of the side effects of the C \texttt{strtok} function. As a simple fix, we updated the system prompt to instruct the LLM to avoid this function, which resolved the issue. Finally, across all of our experiments the LLM only produced a single C model that failed to compile.

\para{3. \tool finds many implementation bugs} To evaluate the quality of tests generated by \tool, we compared the bugs found by prior work with manual models, specifically with SCALE\cite{scale} for DNS and MESSI~\cite{messi} for BGP. Compared to SCALE or MESSI, which used a carefully designed DNS/BGP model constructed manually from RFCs, \tool requires orders of magnitude less modeling effort by leveraging LLMs. On the other hand, we expected the hand-written model to be more effective at revealing bugs.

We present the results from differential testing in \cref{tab:eval-bugs}. The table lists the bugs identified in each of the tested protocol implementations, along with a description of the bug and its type/effect. Additionally, we record whether each bug was found by test cases produced by the prior SCALE or MESSI tool, or whether it would have been found had the SCALE tests been run on the newly added implementations.

In the comparative analysis of DNS implementations, \tool successfully identified a total of \textbf{38} bugs across ten different implementations.  \textbf{26} were unique (\IE, after removing those same bugs that affect multiple implementations) of which \textbf{15} were also found by SCALE and \textbf{11} represent new bugs that SCALE was unable to find. Comparatively, SCALE revealed only \textbf{22} unique bugs, of which \textbf{7} were not found by \tool. \emph{Hence, \tool actually found more bugs than SCALE despite requiring little to no modeling effort.}

Comparing MESSI~\cite{messi} with \tool for BGP protocol implementation testing, \tool replicated 2 bugs in prefix lists and route-filtering found by MESSI. More interestingly, we used \tool to model two BGP features, confederation and route reflection, that were not in the manually written MESSI model and hence were not tested.  Overall, \tool found \textbf{5} unique bugs, of which \textbf{3} were new across all implementations. 

Finally, our experiment with SMTP servers demonstrates how \tool handles stateful protocols. Despite the relative simplicity of the server model, we found \textbf{1} new bug, and with minimal human effort.

\para{4. Additional Insights} We manually examined the seven bugs that SCALE identified but \tool did not. In four of the seven cases, the corresponding test purposely used an invalid zone file, to test how implementations handle them. The \tool-generated DNS models ensure validity, but we could also use \tool to test invalid zone files.  In one case the bug found by SCALE depends on a subtlety in the logic of handling \cname records that the LLM-generated model did not include. Finally, in a few cases that identified a bug went beyond the bounds that we set for the \tool-generated tests, for example involving multiple records in a zone file. 

For BGP confederations, \tool generated an interesting test case where a router $R$ within a confederation had the same sub-AS number as its neighbor $N$'s AS number, where $N$ is outside the confederation. The bug arises when $R$ tries to establish an iBGP connection with $N$ because it thought that $N$ was in the same sub-AS as itself. On the other hand, $N$ thought $R$ was outside its AS, so $N$ tried to initiate an eBGP connection. Klee was probably able to find such a test case because it tends to assign similar values to symbolic variables of the same type unless strictly constrained. 


\section{Related Work}
\label{sec:related-work}

\tool is related to several lines of prior work:

\para{Automated testing}
Work in automatically testing hardware and software  
is often categorized by their 
visibility into the source code. Black box testers~\cite{randoop, snipfuzz, quickcheck, smallcheck, blackbox-sched, pulsar} for instance, generate random tests for implementations whose source code or specification is unavailable. Grey-box testers~\cite{afl, aflfast, angora, aflgo, aflsmart, superion, evosuite} use lightweight program instrumentation to obtain coverage feedback to guide future tests. White-box testers~\cite{sage, klee, dart, whitebox-grammar} use heavyweight symbolic methods with access to source code to directly solve program path constraints. Each approach has limitations -- black and grey-box approaches often get ``stuck'' on complex conditions and can result in low test coverage. Conversely, white-box approaches directly solve for inputs that meet complex conditions but (1) suffer from path explosion due to code complexity, (2) require access to source code, and (3) often require non-trivial source modification to be applied effectively~\cite{dart, sage, protocol-symbex, klee}.

Another approach is model-based or specification-based testing~\cite{quic, quic2, scale, mbt-overview, spec-testing, cedar}. Model-based testing applies to black-box implementations by using a reference model or specification to generate tests. It sidesteps many of the aforementioned issues but requires users to manually craft a model, which is a non-trivial undertaking. 

\para{Protocol testing}
Some testing work has directly targeted protocol implementations due to their importance and complexity~\cite{protocol-testing, protocol-symbex, pulsar, autofuzz, aspfuzz, stateful-fuzz, aflstate, aflnet, scale, messi}. Our approach builds on model-based testing approaches for DNS~\cite{scale} and BGP~\cite{messi}, which combine symbolic execution on the model with differential testing of implementations. However, rather manually creating the protocol models, \tool offloads the vast majority of the modeling task to LLMs through an API for modular and declarative model construction. This approach not only reduces the human burden significantly but also makes \tool useful for testing many protocols, while the prior tools each targets a single protocol.

Other approaches attempt to marry some software engineering testing techniques, such as grey-box fuzzing, with a learned protocol \emph{state} (\EG, the protocol state machine). For instance, Pulsar~\cite{pulsar} and AFLNet~\cite{aflnet} combine black-box and grey-box testing respectively with protocol state machine learning from example traces. Rather than trying to ``reverse engineer'' a protocol from its usage, \tool instead uses an LLM to directly encode a protocol's state and logic.

\para{LLM-based testing}
Recent breakthroughs in natural language processing and program understanding with LLMs have led researchers to reconsider the possibility of automatically testing software using AI models~\cite{llm-fuzzing, llm-fuzzing2, llm-pbt, llm-testing-survey, llm-unit-tests, llm-unit-tests2, llm-mutation-testing, llm-unit-tests3, llm-unit-tests4, llm-conformance-tests}. Most of these works directly ask LLMs to write unit tests for software or fuzz programs (\IE, mutate strings). \tool instead uses LLMs to write simple implementations with well-typed inputs. It then uses off-the-shelf symbolic execution tools to generate exhaustive tests. 
Recently, ChatAFL~\cite{chatafl} enhanced the stateful protocol fuzzer AFLNet~\cite{aflnet} using LLMs. Like \tool, ChatAFL leverages LLM knowledge to improve testing, but it directly modifies the protocol messages that are used as tests. These two approaches are quite distinct and likely have different strengths.

\para{NLP for networks}
Some early work shared the insight that RFCs and other natural language sources could provide useful information for testing network protocols~\cite{nlp-testing} or generating network configurations~\cite{lumi}. In more recent work~\cite{prosper}, the authors used LLMs to extract protocol specifications from RFCs in the form of protocol automata. \tool too extracts a specification, however it does so with human guidance and in the form of a C program that may be combined with symbolic execution.

%
%
%
%

\section{Conclusion}
\label{sec:conclusion}

Despite the excitement, a thorny problem with using LLMs for critical applications is dealing with the errors they make. If LLMs are used to generate code, for instance, we must test~\cite{c_gpt} or verify\cite{clover, hotnets} the output, and iterate if needed.  By contrast, in this paper we argue that if LLMs are used to \emph{test} software, errors in the models and tests produced by the LLM can be finessed using differential testing.  

Specifically, we introduce \emph{model-based testing with LLMs}, a new approach for automatic black-box protocol testing.  We use LLMs for two distinct purposes: to construct protocol models, and also to generate state graphs to find efficient sequences to drive protocols to desired states.
Our approach greatly
reduces the burden of manually creating a 
protocol model, 
applies to black-box implementations, requires no source code changes, and produces tests that find deep functional correctness bugs. 

Our testing framework \tool, based on these principles, found 32 unique bugs in protocol implementations, including 15 new bugs previously undiscovered by manually constructed model-based testing tools. We could quickly build models for BGP features (confederations and route-reflectors) that had not been tested by earlier work (MESSI~\cite{messi}) and also rapidly crafted an SMTP server model from scratch.

Our SMTP experience showed us that LLMs can also be used to drive protocols to specified states for testing, but we have only scratched the surface. We hope to explore this capability further to automatically test more complex stateful protocols like TCP and HTTP.

\section*{Acknowledgments}
We thank the developers of various open-source implementations for their feedback on our bug reports. The authors, Rajdeep Mondal, Rathin Singha, Todd Millstein and George Varghese, were partially supported by NSF Grant 2402958.

\newpage
\bibliographystyle{ACM-Reference-Format}
\bibliography{paper}


\begin{thebibliography}{85}


\ifx \showCODEN    \undefined \def \showCODEN     #1{\unskip}     \fi
\ifx \showDOI      \undefined \def \showDOI       #1{#1}\fi
\ifx \showISBNx    \undefined \def \showISBNx     #1{\unskip}     \fi
\ifx \showISBNxiii \undefined \def \showISBNxiii  #1{\unskip}     \fi
\ifx \showISSN     \undefined \def \showISSN      #1{\unskip}     \fi
\ifx \showLCCN     \undefined \def \showLCCN      #1{\unskip}     \fi
\ifx \shownote     \undefined \def \shownote      #1{#1}          \fi
\ifx \showarticletitle \undefined \def \showarticletitle #1{#1}   \fi
\ifx \showURL      \undefined \def \showURL       {\relax}        \fi
\providecommand\bibfield[2]{#2}
\providecommand\bibinfo[2]{#2}
\providecommand\natexlab[1]{#1}
\providecommand\showeprint[2][]{arXiv:#2}

\bibitem[\protect\citeauthoryear{??}{ope}{2024}]%
        {opensmtpd}
 \bibinfo{year}{2024}\natexlab{}.
\newblock \bibinfo{title}{OpenSMTPD}.
\newblock \bibinfo{howpublished}{\\\url{https://www.opensmtpd.org/}}.
  (\bibinfo{year}{2024}).
\newblock


\bibitem[\protect\citeauthoryear{aiosmtpd community}{aiosmtpd
  community}{2024}]%
        {aiosmtpd}
\bibfield{author}{\bibinfo{person}{aiosmtpd community}.}
  \bibinfo{year}{2024}\natexlab{}.
\newblock \bibinfo{title}{aiosmtpd - An asyncio based SMTP server}.
\newblock
  \bibinfo{howpublished}{\\\url{https://aiosmtpd.aio-libs.org/en/latest/}}.
  (\bibinfo{year}{2024}).
\newblock


\bibitem[\protect\citeauthoryear{Ba, B{\"o}hme, Mirzamomen, and
  Roychoudhury}{Ba et~al\mbox{.}}{2022}]%
        {stateful-fuzz}
\bibfield{author}{\bibinfo{person}{Jinsheng Ba}, \bibinfo{person}{Marcel
  B{\"o}hme}, \bibinfo{person}{Zahra Mirzamomen}, {and} \bibinfo{person}{Abhik
  Roychoudhury}.} \bibinfo{year}{2022}\natexlab{}.
\newblock \showarticletitle{Stateful greybox fuzzing}. In
  \bibinfo{booktitle}{{\em 31st USENIX Security Symposium (USENIX Security
  22)}}. \bibinfo{pages}{3255--3272}.
\newblock


\bibitem[\protect\citeauthoryear{Berinato}{Berinato}{2023}]%
        {outage-medical}
\bibfield{author}{\bibinfo{person}{Scott Berinato}.}
  \bibinfo{year}{2023}\natexlab{}.
\newblock \bibinfo{title}{All systems down}.
\newblock
  \bibinfo{howpublished}{\url{https://www.computerworld.com/article/2581420/all-systems-down.html}}.
    (\bibinfo{year}{2023}).
\newblock
\newblock
\shownote{Accessed: 2023-9-29.}


\bibitem[\protect\citeauthoryear{Black and Community}{Black and
  Community}{2023}]%
        {gdnsd}
\bibfield{author}{\bibinfo{person}{Brandon~L Black} {and}
  \bibinfo{person}{Community}.} \bibinfo{year}{2023}\natexlab{}.
\newblock \bibinfo{title}{GDNSD}.
\newblock \bibinfo{howpublished}{\\\url{https://gdnsd.org/}}.
  (\bibinfo{year}{2023}).
\newblock
\newblock
\shownote{\\Eywa commit:
  \url{https://github.com/gdnsd/gdnsd/tree/877e15cf55593fa618d2009027e928d5f52da775}.}


\bibitem[\protect\citeauthoryear{Bochmann and Petrenko}{Bochmann and
  Petrenko}{1994}]%
        {protocol-testing}
\bibfield{author}{\bibinfo{person}{Gregor~V Bochmann} {and}
  \bibinfo{person}{Alexandre Petrenko}.} \bibinfo{year}{1994}\natexlab{}.
\newblock \showarticletitle{Protocol testing: review of methods and relevance
  for software testing}. In \bibinfo{booktitle}{{\em Proceedings of the 1994
  ACM SIGSOFT international symposium on Software testing and analysis}}.
  \bibinfo{pages}{109--124}.
\newblock


\bibitem[\protect\citeauthoryear{B{\"o}hme, Pham, Nguyen, and
  Roychoudhury}{B{\"o}hme et~al\mbox{.}}{2017}]%
        {aflgo}
\bibfield{author}{\bibinfo{person}{Marcel B{\"o}hme},
  \bibinfo{person}{Van-Thuan Pham}, \bibinfo{person}{Manh-Dung Nguyen}, {and}
  \bibinfo{person}{Abhik Roychoudhury}.} \bibinfo{year}{2017}\natexlab{}.
\newblock \showarticletitle{Directed greybox fuzzing}. In
  \bibinfo{booktitle}{{\em Proceedings of the 2017 ACM SIGSAC conference on
  computer and communications security}}. \bibinfo{pages}{2329--2344}.
\newblock


\bibitem[\protect\citeauthoryear{B{\"o}hme, Pham, and Roychoudhury}{B{\"o}hme
  et~al\mbox{.}}{2016}]%
        {aflfast}
\bibfield{author}{\bibinfo{person}{Marcel B{\"o}hme},
  \bibinfo{person}{Van-Thuan Pham}, {and} \bibinfo{person}{Abhik
  Roychoudhury}.} \bibinfo{year}{2016}\natexlab{}.
\newblock \showarticletitle{Coverage-based greybox fuzzing as markov chain}. In
  \bibinfo{booktitle}{{\em Proceedings of the 2016 ACM SIGSAC Conference on
  Computer and Communications Security}}. \bibinfo{pages}{1032--1043}.
\newblock


\bibitem[\protect\citeauthoryear{Brook}{Brook}{2023}]%
        {bug-cisco}
\bibfield{author}{\bibinfo{person}{Chris Brook}.}
  \bibinfo{year}{2023}\natexlab{}.
\newblock \bibinfo{title}{Cisco Fixes DoS, Authentication Bypass
  Vulnerabilities, OSPF Bug}.
\newblock
  \bibinfo{howpublished}{\url{https://threatpost.com/cisco-fixes-dos-authentication-bypass-vulnerabilities-ospf-bug/127185/}}.
    (\bibinfo{year}{2023}).
\newblock
\newblock
\shownote{Accessed: 2023-9-29.}


\bibitem[\protect\citeauthoryear{Cadar, Dunbar, Engler, et~al\mbox{.}}{Cadar
  et~al\mbox{.}}{2008}]%
        {klee}
\bibfield{author}{\bibinfo{person}{Cristian Cadar}, \bibinfo{person}{Daniel
  Dunbar}, \bibinfo{person}{Dawson~R Engler}, {et~al\mbox{.}}}
  \bibinfo{year}{2008}\natexlab{}.
\newblock \showarticletitle{Klee: Unassisted and automatic generation of
  high-coverage tests for complex systems programs.}. In
  \bibinfo{booktitle}{{\em OSDI}}, Vol.~\bibinfo{volume}{8}.
  \bibinfo{pages}{209--224}.
\newblock


\bibitem[\protect\citeauthoryear{Chen and Chen}{Chen and Chen}{2018}]%
        {angora}
\bibfield{author}{\bibinfo{person}{Peng Chen} {and} \bibinfo{person}{Hao
  Chen}.} \bibinfo{year}{2018}\natexlab{}.
\newblock \showarticletitle{Angora: Efficient fuzzing by principled search}. In
  \bibinfo{booktitle}{{\em 2018 IEEE Symposium on Security and Privacy (SP)}}.
  IEEE, \bibinfo{pages}{711--725}.
\newblock


\bibitem[\protect\citeauthoryear{Claessen and Hughes}{Claessen and
  Hughes}{2000}]%
        {quickcheck}
\bibfield{author}{\bibinfo{person}{Koen Claessen} {and} \bibinfo{person}{John
  Hughes}.} \bibinfo{year}{2000}\natexlab{}.
\newblock \showarticletitle{QuickCheck: a lightweight tool for random testing
  of Haskell programs}. In \bibinfo{booktitle}{{\em Proceedings of the fifth
  ACM SIGPLAN international conference on Functional programming}}.
  \bibinfo{pages}{268--279}.
\newblock


\bibitem[\protect\citeauthoryear{community}{community}{2023}]%
        {coredns}
\bibfield{author}{\bibinfo{person}{CoreDNS community}.}
  \bibinfo{year}{2023}\natexlab{}.
\newblock \bibinfo{title}{CoreDNS}.
\newblock \bibinfo{howpublished}{\\\url{https://coredns.io/}}.
  (\bibinfo{year}{2023}).
\newblock
\newblock
\shownote{\\Eywa commit:
  \url{https://github.com/coredns/coredns/tree/45923b6e12a2eabaf55d7380e6df4e7354a1207}\\
  SCALE commit:
  \url{https://github.com/coredns/coredns/tree/6edc8fe7f6c2f57844c8ee7f7f5deef71085ebe8}.}


\bibitem[\protect\citeauthoryear{Community}{Community}{2023a}]%
        {dnspython}
\bibfield{author}{\bibinfo{person}{Dnspython Community}.}
  \bibinfo{year}{2023}\natexlab{a}.
\newblock \bibinfo{title}{Dnspython}.
\newblock
  \bibinfo{howpublished}{\\\url{https://dnspython.readthedocs.io/en/latest/index.html}}.
    (\bibinfo{year}{2023}).
\newblock


\bibitem[\protect\citeauthoryear{community}{community}{2024a}]%
        {exabgp}
\bibfield{author}{\bibinfo{person}{ExaBGP community}.}
  \bibinfo{year}{2024}\natexlab{a}.
\newblock \bibinfo{title}{ExaBGP}.
\newblock
  \bibinfo{howpublished}{\\\url{https://github.com/Exa-Networks/exabgp}}.
  (\bibinfo{year}{2024}).
\newblock


\bibitem[\protect\citeauthoryear{community}{community}{2024b}]%
        {frr}
\bibfield{author}{\bibinfo{person}{FRR community}.}
  \bibinfo{year}{2024}\natexlab{b}.
\newblock \bibinfo{title}{The FRRouting Protocol Suite}.
\newblock \bibinfo{howpublished}{\\\url{https://frrouting.org/}}.
  (\bibinfo{year}{2024}).
\newblock
\newblock
\shownote{\\Version:
  \url{https://github.com/FRRouting/frr/releases/tag/frr-10.1.2}.}


\bibitem[\protect\citeauthoryear{community}{community}{2024c}]%
        {gobgp}
\bibfield{author}{\bibinfo{person}{GoBGP community}.}
  \bibinfo{year}{2024}\natexlab{c}.
\newblock \bibinfo{title}{GoBGP}.
\newblock \bibinfo{howpublished}{\\\url{https://github.com/osrg/gobgp}}.
  (\bibinfo{year}{2024}).
\newblock


\bibitem[\protect\citeauthoryear{Community}{Community}{2023b}]%
        {PowerDNS}
\bibfield{author}{\bibinfo{person}{PowerDNS Community}.}
  \bibinfo{year}{2023}\natexlab{b}.
\newblock \bibinfo{title}{PowerDNS}.
\newblock \bibinfo{howpublished}{\\\url{https://www.powerdns.com/}}.
  (\bibinfo{year}{2023}).
\newblock
\newblock
\shownote{\\Eywa commit:
  \url{https://github.com/PowerDNS/pdns/tree/8314f12e92a8b75e33438bc7c16c6430028fbef9}\\
  SCALE commit:
  \url{https://github.com/PowerDNS/pdns/tree/a03aaad7554483ee6efe72a81eda00a9d1a94fe5}.}


\bibitem[\protect\citeauthoryear{Consortium}{Consortium}{2023}]%
        {bind9}
\bibfield{author}{\bibinfo{person}{Internet~Systems Consortium}.}
  \bibinfo{year}{2023}\natexlab{}.
\newblock \bibinfo{title}{BIND 9}.
\newblock \bibinfo{howpublished}{\\\url{https://www.isc.org/bind/}}.
  (\bibinfo{year}{2023}).
\newblock
\newblock
\shownote{\\Eywa commit:
  \url{https://gitlab.isc.org/isc-projects/bind9/-/tree/85ee12f60edb6b79535f6f226250ac471d68fbab}\\
  SCALE commit:
  \url{https://gitlab.isc.org/isc-projects/bind9/-/tree/dbcf683c1a57f49876e329fca183cb39d20ca3a4}.}


\bibitem[\protect\citeauthoryear{curl}{curl}{2023a}]%
        {bug-ftp2}
\bibfield{author}{\bibinfo{person}{curl}.} \bibinfo{year}{2023}\natexlab{a}.
\newblock \bibinfo{title}{FTP Server Response Buffer Overflow}.
\newblock
  \bibinfo{howpublished}{\url{https://curl.se/docs/CVE-2000-0973.html}}.
  (\bibinfo{year}{2023}).
\newblock
\newblock
\shownote{Accessed: 2023-9-29.}


\bibitem[\protect\citeauthoryear{curl}{curl}{2023b}]%
        {bug-ftp1}
\bibfield{author}{\bibinfo{person}{curl}.} \bibinfo{year}{2023}\natexlab{b}.
\newblock \bibinfo{title}{FTP shutdown response buffer overflow}.
\newblock
  \bibinfo{howpublished}{\url{https://curl.se/docs/CVE-2018-1000300.html}}.
  (\bibinfo{year}{2023}).
\newblock
\newblock
\shownote{Accessed: 2023-9-29.}


\bibitem[\protect\citeauthoryear{cyberstanc}{cyberstanc}{2023}]%
        {bug-icmp}
\bibfield{author}{\bibinfo{person}{cyberstanc}.}
  \bibinfo{year}{2023}\natexlab{}.
\newblock \bibinfo{title}{Pinging our way to Remote Code Execution: The New
  ICMP Vulnerability You Need to Know About!}
\newblock
  \bibinfo{howpublished}{\url{https://cyberstanc.com/blog/pinging-our-way-to-remote-code-execution-the-new-icmp-vulnerability-you-need-to-know-about/}}.
    (\bibinfo{year}{2023}).
\newblock
\newblock
\shownote{Accessed: 2023-9-29.}


\bibitem[\protect\citeauthoryear{CZ.NIC}{CZ.NIC}{2023}]%
        {knot}
\bibfield{author}{\bibinfo{person}{CZ.NIC}.} \bibinfo{year}{2023}\natexlab{}.
\newblock \bibinfo{title}{Knot}.
\newblock \bibinfo{howpublished}{\\\url{https://www.knot-dns.cz/}}.
  (\bibinfo{year}{2023}).
\newblock
\newblock
\shownote{\\Eywa commit:
  \url{https://gitlab.nic.cz/knot/knot-dns/-/tree/c08e5738b6eed43b052a127d56db6451106386fa}\\
  SCALE commit:
  \url{https://gitlab.nic.cz/knot/knot-dns/-/tree/89aaeb729a0856fefaed111c114ebb8a5a3f4ed2}.}


\bibitem[\protect\citeauthoryear{developer community}{developer
  community}{2024}]%
        {smtpd}
\bibfield{author}{\bibinfo{person}{Python developer community}.}
  \bibinfo{year}{2024}\natexlab{}.
\newblock \bibinfo{title}{SMTPD Python library}.
\newblock
  \bibinfo{howpublished}{\\\url{https://docs.python.org/3.10/library/smtpd.html}}.
    (\bibinfo{year}{2024}).
\newblock


\bibitem[\protect\citeauthoryear{Disselkoen, Eline, He, Headley, Hicks,
  Hietala, Kastner, Mamat, McCutchen, Rungta, Shah, Torlak, and
  Wells}{Disselkoen et~al\mbox{.}}{2024}]%
        {cedar}
\bibfield{author}{\bibinfo{person}{Craig Disselkoen}, \bibinfo{person}{Aaron
  Eline}, \bibinfo{person}{Shaobo He}, \bibinfo{person}{Kyle Headley},
  \bibinfo{person}{Michael Hicks}, \bibinfo{person}{Kesha Hietala},
  \bibinfo{person}{John Kastner}, \bibinfo{person}{Anwar Mamat},
  \bibinfo{person}{Matt McCutchen}, \bibinfo{person}{Neha Rungta},
  \bibinfo{person}{Bhakti Shah}, \bibinfo{person}{Emina Torlak}, {and}
  \bibinfo{person}{Andrew Wells}.} \bibinfo{year}{2024}\natexlab{}.
\newblock \showarticletitle{How We Built Cedar: A Verification-Guided
  Approach}. In \bibinfo{booktitle}{{\em Companion Proceedings of the 32nd ACM
  International Conference on the Foundations of Software Engineering}} {\em
  (\bibinfo{series}{FSE 2024})}. \bibinfo{publisher}{Association for Computing
  Machinery}, \bibinfo{address}{New York, NY, USA}, \bibinfo{pages}{351–357}.
\newblock
\showISBNx{9798400706585}
\showDOI{%
\url{https://doi.org/10.1145/3663529.3663854}}


\bibitem[\protect\citeauthoryear{Duffy}{Duffy}{2023}]%
        {bug-juniper}
\bibfield{author}{\bibinfo{person}{Jim Duffy}.}
  \bibinfo{year}{2023}\natexlab{}.
\newblock \bibinfo{title}{BGP bug bites Juniper software}.
\newblock
  \bibinfo{howpublished}{\url{https://www.networkworld.com/article/2289950/bgp-bug-bites-juniper-software.html}}.
    (\bibinfo{year}{2023}).
\newblock
\newblock
\shownote{Accessed: 2023-9-29.}


\bibitem[\protect\citeauthoryear{Dutta}{Dutta}{2023}]%
        {outage-bgp}
\bibfield{author}{\bibinfo{person}{Tushar~Subhra Dutta}.}
  \bibinfo{year}{2023}\natexlab{}.
\newblock \bibinfo{title}{BGP Error Handling Flaw Leads to Prolonged Network
  Outage}.
\newblock
  \bibinfo{howpublished}{\url{https://cybersecuritynews.com/bgp-error-handling-flaw/}}.
    (\bibinfo{year}{2023}).
\newblock
\newblock
\shownote{Accessed: 2023-9-29.}


\bibitem[\protect\citeauthoryear{EURid.eu}{EURid.eu}{2023}]%
        {yadifa}
\bibfield{author}{\bibinfo{person}{EURid.eu}.} \bibinfo{year}{2023}\natexlab{}.
\newblock \bibinfo{title}{YADIFA}.
\newblock \bibinfo{howpublished}{\\\url{https://www.yadifa.eu/}}.
  (\bibinfo{year}{2023}).
\newblock
\newblock
\shownote{\\Eywa commit:
  \url{https://github.com/yadifa/yadifa/tree/9bb6facead9e7ba222962b2980f85fa6ba02e465}\\
  SCALE commit:
  \url{https://github.com/yadifa/yadifa/tree/dc5bed2fb8ec204af9b65eeb91934c2c85098cbb}.}


\bibitem[\protect\citeauthoryear{Feng, Sun, Zhu, Xue, Wen, Liu, Nepal, and
  Xiang}{Feng et~al\mbox{.}}{2021}]%
        {snipfuzz}
\bibfield{author}{\bibinfo{person}{Xiaotao Feng}, \bibinfo{person}{Ruoxi Sun},
  \bibinfo{person}{Xiaogang Zhu}, \bibinfo{person}{Minhui Xue},
  \bibinfo{person}{Sheng Wen}, \bibinfo{person}{Dongxi Liu},
  \bibinfo{person}{Surya Nepal}, {and} \bibinfo{person}{Yang Xiang}.}
  \bibinfo{year}{2021}\natexlab{}.
\newblock \showarticletitle{Snipuzz: Black-box fuzzing of iot firmware via
  message snippet inference}. In \bibinfo{booktitle}{{\em Proceedings of the
  2021 ACM SIGSAC Conference on Computer and Communications Security}}.
  \bibinfo{pages}{337--350}.
\newblock


\bibitem[\protect\citeauthoryear{Fogel, Fung, Pedrosa, Walraed-Sullivan,
  Govindan, Mahajan, and Millstein}{Fogel et~al\mbox{.}}{2015}]%
        {batfish}
\bibfield{author}{\bibinfo{person}{Ari Fogel}, \bibinfo{person}{Stanley Fung},
  \bibinfo{person}{Luis Pedrosa}, \bibinfo{person}{Meg Walraed-Sullivan},
  \bibinfo{person}{Ramesh Govindan}, \bibinfo{person}{Ratul Mahajan}, {and}
  \bibinfo{person}{Todd Millstein}.} \bibinfo{year}{2015}\natexlab{}.
\newblock \showarticletitle{A general approach to network configuration
  analysis}. In \bibinfo{booktitle}{{\em Proceedings of the 12th USENIX
  Conference on Networked Systems Design and Implementation}} {\em
  (\bibinfo{series}{NSDI'15})}. \bibinfo{publisher}{USENIX Association},
  \bibinfo{address}{USA}, \bibinfo{pages}{469–483}.
\newblock
\showISBNx{9781931971218}


\bibitem[\protect\citeauthoryear{Fraser and Arcuri}{Fraser and Arcuri}{2011}]%
        {evosuite}
\bibfield{author}{\bibinfo{person}{Gordon Fraser} {and} \bibinfo{person}{Andrea
  Arcuri}.} \bibinfo{year}{2011}\natexlab{}.
\newblock \showarticletitle{Evosuite: automatic test suite generation for
  object-oriented software}. In \bibinfo{booktitle}{{\em Proceedings of the
  19th ACM SIGSOFT symposium and the 13th European conference on Foundations of
  software engineering}}. \bibinfo{pages}{416--419}.
\newblock


\bibitem[\protect\citeauthoryear{Fry and Community}{Fry and Community}{2023}]%
        {trustdns}
\bibfield{author}{\bibinfo{person}{Benjamin Fry} {and}
  \bibinfo{person}{Community}.} \bibinfo{year}{2023}\natexlab{}.
\newblock \bibinfo{title}{Hickory-DNS}.
\newblock
  \bibinfo{howpublished}{\\\url{https://github.com/hickory-dns/hickory-dns}}.
  (\bibinfo{year}{2023}).
\newblock
\newblock
\shownote{\\Eywa commit:
  \url{https://github.com/hickory-dns/hickory-dns/tree/65c5327ef6b8dbda92654837b8b5cb31fa0000ad}\\
  SCALE commit:
  \url{https://github.com/bluejekyll/trust-dns/tree/7d9b186121fb5cb331cf2ec6baa47846b83de8fc}.}


\bibitem[\protect\citeauthoryear{Gascon, Wressnegger, Yamaguchi, Arp, and
  Rieck}{Gascon et~al\mbox{.}}{2015}]%
        {pulsar}
\bibfield{author}{\bibinfo{person}{Hugo Gascon}, \bibinfo{person}{Christian
  Wressnegger}, \bibinfo{person}{Fabian Yamaguchi}, \bibinfo{person}{Daniel
  Arp}, {and} \bibinfo{person}{Konrad Rieck}.} \bibinfo{year}{2015}\natexlab{}.
\newblock \showarticletitle{Pulsar: Stateful black-box fuzzing of proprietary
  network protocols}. In \bibinfo{booktitle}{{\em Security and Privacy in
  Communication Networks: 11th EAI International Conference, SecureComm 2015,
  Dallas, TX, USA, October 26-29, 2015, Proceedings 11}}. Springer,
  \bibinfo{pages}{330--347}.
\newblock


\bibitem[\protect\citeauthoryear{github}{github}{2023}]%
        {afl}
\bibfield{author}{\bibinfo{person}{github}.} \bibinfo{year}{2023}\natexlab{}.
\newblock \bibinfo{title}{google/AFL}.
\newblock \bibinfo{howpublished}{\url{https://github.com/google/AFL}}.
  (\bibinfo{year}{2023}).
\newblock
\newblock
\shownote{Accessed: 2023-9-29.}


\bibitem[\protect\citeauthoryear{Godefroid, Kiezun, and Levin}{Godefroid
  et~al\mbox{.}}{2008}]%
        {whitebox-grammar}
\bibfield{author}{\bibinfo{person}{Patrice Godefroid}, \bibinfo{person}{Adam
  Kiezun}, {and} \bibinfo{person}{Michael~Y Levin}.}
  \bibinfo{year}{2008}\natexlab{}.
\newblock \showarticletitle{Grammar-based whitebox fuzzing}. In
  \bibinfo{booktitle}{{\em Proceedings of the 29th ACM SIGPLAN conference on
  programming language design and implementation}}. \bibinfo{pages}{206--215}.
\newblock


\bibitem[\protect\citeauthoryear{Godefroid, Klarlund, and Sen}{Godefroid
  et~al\mbox{.}}{2005}]%
        {dart}
\bibfield{author}{\bibinfo{person}{Patrice Godefroid}, \bibinfo{person}{Nils
  Klarlund}, {and} \bibinfo{person}{Koushik Sen}.}
  \bibinfo{year}{2005}\natexlab{}.
\newblock \showarticletitle{DART: Directed automated random testing}. In
  \bibinfo{booktitle}{{\em Proceedings of the 2005 ACM SIGPLAN conference on
  Programming language design and implementation}}. \bibinfo{pages}{213--223}.
\newblock


\bibitem[\protect\citeauthoryear{Godefroid, Levin, and Molnar}{Godefroid
  et~al\mbox{.}}{2012}]%
        {sage}
\bibfield{author}{\bibinfo{person}{Patrice Godefroid},
  \bibinfo{person}{Michael~Y Levin}, {and} \bibinfo{person}{David Molnar}.}
  \bibinfo{year}{2012}\natexlab{}.
\newblock \showarticletitle{SAGE: whitebox fuzzing for security testing}.
\newblock \bibinfo{journal}{{\it Commun. ACM}} \bibinfo{volume}{55},
  \bibinfo{number}{3} (\bibinfo{year}{2012}), \bibinfo{pages}{40--44}.
\newblock


\bibitem[\protect\citeauthoryear{Gorbunov and Rosenbloom}{Gorbunov and
  Rosenbloom}{2010}]%
        {autofuzz}
\bibfield{author}{\bibinfo{person}{Serge Gorbunov} {and}
  \bibinfo{person}{Arnold Rosenbloom}.} \bibinfo{year}{2010}\natexlab{}.
\newblock \showarticletitle{Autofuzz: Automated network protocol fuzzing
  framework}.
\newblock \bibinfo{journal}{{\em Ijcsns\/}} \bibinfo{volume}{10},
  \bibinfo{number}{8} (\bibinfo{year}{2010}), \bibinfo{pages}{239}.
\newblock


\bibitem[\protect\citeauthoryear{Hu, Zhang, and Yin}{Hu et~al\mbox{.}}{2023}]%
        {llm-fuzzing2}
\bibfield{author}{\bibinfo{person}{Jie Hu}, \bibinfo{person}{Qian Zhang}, {and}
  \bibinfo{person}{Heng Yin}.} \bibinfo{year}{2023}\natexlab{}.
\newblock \showarticletitle{Augmenting Greybox Fuzzing with Generative AI}.
\newblock \bibinfo{journal}{{\em arXiv preprint arXiv:2306.06782\/}}
  (\bibinfo{year}{2023}).
\newblock


\bibitem[\protect\citeauthoryear{ipSpace}{ipSpace}{2023}]%
        {bug-ios}
\bibfield{author}{\bibinfo{person}{ipSpace}.} \bibinfo{year}{2023}\natexlab{}.
\newblock \bibinfo{title}{Oversized AS Paths: Cisco IOS Bug Details}.
\newblock
  \bibinfo{howpublished}{\url{https://blog.ipspace.net/2009/02/oversized-as-paths-cisco-ios-bug.html}}.
    (\bibinfo{year}{2023}).
\newblock
\newblock
\shownote{Accessed: 2023-9-29.}


\bibitem[\protect\citeauthoryear{Jacobs, Pfitscher, Ribeiro, Ferreira,
  Granville, Willinger, and Rao}{Jacobs et~al\mbox{.}}{2021}]%
        {lumi}
\bibfield{author}{\bibinfo{person}{Arthur~S Jacobs}, \bibinfo{person}{Ricardo~J
  Pfitscher}, \bibinfo{person}{Rafael~H Ribeiro}, \bibinfo{person}{Ronaldo~A
  Ferreira}, \bibinfo{person}{Lisandro~Z Granville}, \bibinfo{person}{Walter
  Willinger}, {and} \bibinfo{person}{Sanjay~G Rao}.}
  \bibinfo{year}{2021}\natexlab{}.
\newblock \showarticletitle{Hey, lumi! using natural language for
  $\{$intent-based$\}$ network management}. In \bibinfo{booktitle}{{\em 2021
  USENIX Annual Technical Conference (USENIX ATC 21)}}.
  \bibinfo{pages}{625--639}.
\newblock


\bibitem[\protect\citeauthoryear{Jacobson}{Jacobson}{1988}]%
        {congestion-avoidance}
\bibfield{author}{\bibinfo{person}{Van Jacobson}.}
  \bibinfo{year}{1988}\natexlab{}.
\newblock \showarticletitle{Congestion avoidance and control}.
\newblock \bibinfo{journal}{{\em ACM SIGCOMM computer communication review\/}}
  \bibinfo{volume}{18}, \bibinfo{number}{4} (\bibinfo{year}{1988}),
  \bibinfo{pages}{314--329}.
\newblock


\bibitem[\protect\citeauthoryear{Johansson, Svensson, Larson, Almgren, and
  Gulisano}{Johansson et~al\mbox{.}}{2014}]%
        {tfuzz}
\bibfield{author}{\bibinfo{person}{William Johansson}, \bibinfo{person}{Martin
  Svensson}, \bibinfo{person}{Ulf~E Larson}, \bibinfo{person}{Magnus Almgren},
  {and} \bibinfo{person}{Vincenzo Gulisano}.} \bibinfo{year}{2014}\natexlab{}.
\newblock \showarticletitle{T-Fuzz: Model-based fuzzing for robustness testing
  of telecommunication protocols}. In \bibinfo{booktitle}{{\em 2014 IEEE
  Seventh International Conference on Software Testing, Verification and
  Validation}}. IEEE, \bibinfo{pages}{323--332}.
\newblock


\bibitem[\protect\citeauthoryear{Kakarla, Beckett, Arzani, Millstein, and
  Varghese}{Kakarla et~al\mbox{.}}{2020}]%
        {groot}
\bibfield{author}{\bibinfo{person}{Siva Kesava~Reddy Kakarla},
  \bibinfo{person}{Ryan Beckett}, \bibinfo{person}{Behnaz Arzani},
  \bibinfo{person}{Todd Millstein}, {and} \bibinfo{person}{George Varghese}.}
  \bibinfo{year}{2020}\natexlab{}.
\newblock \showarticletitle{GRooT: Proactive Verification of DNS
  Configurations}. In \bibinfo{booktitle}{{\em Proceedings of the Annual
  Conference of the ACM Special Interest Group on Data Communication on the
  Applications, Technologies, Architectures, and Protocols for Computer
  Communication}} {\em (\bibinfo{series}{SIGCOMM '20})}.
  \bibinfo{publisher}{Association for Computing Machinery},
  \bibinfo{address}{New York, NY, USA}, \bibinfo{pages}{310–328}.
\newblock
\showISBNx{9781450379557}
\showDOI{%
\url{https://doi.org/10.1145/3387514.3405871}}


\bibitem[\protect\citeauthoryear{Kakarla, Beckett, Millstein, and
  Varghese}{Kakarla et~al\mbox{.}}{2022}]%
        {scale}
\bibfield{author}{\bibinfo{person}{Siva Kesava~Reddy Kakarla},
  \bibinfo{person}{Ryan Beckett}, \bibinfo{person}{Todd Millstein}, {and}
  \bibinfo{person}{George Varghese}.} \bibinfo{year}{2022}\natexlab{}.
\newblock \showarticletitle{{SCALE}: Automatically Finding {RFC} Compliance
  Bugs in {DNS} Nameservers}. In \bibinfo{booktitle}{{\em 19th USENIX Symposium
  on Networked Systems Design and Implementation (NSDI 22)}}.
  \bibinfo{publisher}{USENIX Association}, \bibinfo{address}{Renton, WA},
  \bibinfo{pages}{307--323}.
\newblock
\showISBNx{978-1-939133-27-4}
\showURL{%
\url{https://www.usenix.org/conference/nsdi22/presentation/kakarla}}


\bibitem[\protect\citeauthoryear{Khanfir, Degiovanni, Papadakis, and
  Traon}{Khanfir et~al\mbox{.}}{2023}]%
        {llm-mutation-testing}
\bibfield{author}{\bibinfo{person}{Ahmed Khanfir}, \bibinfo{person}{Renzo
  Degiovanni}, \bibinfo{person}{Mike Papadakis}, {and} \bibinfo{person}{Yves~Le
  Traon}.} \bibinfo{year}{2023}\natexlab{}.
\newblock \showarticletitle{Efficient Mutation Testing via Pre-Trained Language
  Models}.
\newblock \bibinfo{journal}{{\em arXiv preprint arXiv:2301.03543\/}}
  (\bibinfo{year}{2023}).
\newblock


\bibitem[\protect\citeauthoryear{Khurshid and Marinov}{Khurshid and
  Marinov}{2004}]%
        {spec-testing}
\bibfield{author}{\bibinfo{person}{Sarfraz Khurshid} {and}
  \bibinfo{person}{Darko Marinov}.} \bibinfo{year}{2004}\natexlab{}.
\newblock \showarticletitle{TestEra: Specification-based testing of Java
  programs using SAT}.
\newblock \bibinfo{journal}{{\em Automated Software Engineering\/}}
  \bibinfo{volume}{11} (\bibinfo{year}{2004}), \bibinfo{pages}{403--434}.
\newblock


\bibitem[\protect\citeauthoryear{Kitagawa, Hanaoka, and Kono}{Kitagawa
  et~al\mbox{.}}{2010}]%
        {aspfuzz}
\bibfield{author}{\bibinfo{person}{Takahisa Kitagawa}, \bibinfo{person}{Miyuki
  Hanaoka}, {and} \bibinfo{person}{Kenji Kono}.}
  \bibinfo{year}{2010}\natexlab{}.
\newblock \showarticletitle{Aspfuzz: A state-aware protocol fuzzer based on
  application-layer protocols}. In \bibinfo{booktitle}{{\em The IEEE symposium
  on Computers and Communications}}. IEEE, \bibinfo{pages}{202--208}.
\newblock


\bibitem[\protect\citeauthoryear{Labs}{Labs}{2023a}]%
        {nsd}
\bibfield{author}{\bibinfo{person}{NLnet Labs}.}
  \bibinfo{year}{2023}\natexlab{a}.
\newblock \bibinfo{title}{NSD}.
\newblock
  \bibinfo{howpublished}{\\\url{https://nlnetlabs.nl/projects/nsd/about/}}.
  (\bibinfo{year}{2023}).
\newblock
\newblock
\shownote{\\Eywa commit:
  \url{https://github.com/NLnetLabs/nsd/tree/42208cc79ebd9c8594f15ef859c2fa426851ca9d}\\
  SCALE commit:
  \url{https://github.com/NLnetLabs/nsd/tree/4043a5ab7be7abaec969011e48e4d0d60a0056a6}.}


\bibitem[\protect\citeauthoryear{Labs}{Labs}{2023b}]%
        {twisted}
\bibfield{author}{\bibinfo{person}{Twisted~Matrix Labs}.}
  \bibinfo{year}{2023}\natexlab{b}.
\newblock \bibinfo{title}{TwistedNames}.
\newblock \bibinfo{howpublished}{\\\url{https://twisted.org/}}.
  (\bibinfo{year}{2023}).
\newblock
\newblock
\shownote{\\Eywa commit:
  \url{https://github.com/twisted/twisted/tree/157cd8e659705940e895d321339d467e76ae9d0a}.}


\bibitem[\protect\citeauthoryear{Liu, Xia, Wang, and Zhang}{Liu
  et~al\mbox{.}}{2023b}]%
        {c_gpt}
\bibfield{author}{\bibinfo{person}{Jiawei Liu}, \bibinfo{person}{Chunqiu~Steven
  Xia}, \bibinfo{person}{Yuyao Wang}, {and} \bibinfo{person}{Lingming Zhang}.}
  \bibinfo{year}{2023}\natexlab{b}.
\newblock \showarticletitle{Is your code generated by ChatGPT really correct?
  rigorous evaluation of large language models for code generation}. In
  \bibinfo{booktitle}{{\em Proceedings of the 37th International Conference on
  Neural Information Processing Systems}} {\em (\bibinfo{series}{NIPS '23})}.
  \bibinfo{publisher}{Curran Associates Inc.}, \bibinfo{address}{Red Hook, NY,
  USA}, Article \bibinfo{articleno}{943}, \bibinfo{numpages}{15}~pages.
\newblock


\bibitem[\protect\citeauthoryear{Liu, Duan, Heimes, Bearzi, Vieli, Basin, and
  Perrig}{Liu et~al\mbox{.}}{2023a}]%
        {dns_maude}
\bibfield{author}{\bibinfo{person}{Si Liu}, \bibinfo{person}{Huayi Duan},
  \bibinfo{person}{Lukas Heimes}, \bibinfo{person}{Marco Bearzi},
  \bibinfo{person}{Jodok Vieli}, \bibinfo{person}{David Basin}, {and}
  \bibinfo{person}{Adrian Perrig}.} \bibinfo{year}{2023}\natexlab{a}.
\newblock \showarticletitle{A Formal Framework for End-to-End DNS Resolution}.
  In \bibinfo{booktitle}{{\em Proceedings of the ACM SIGCOMM 2023 Conference}}
  {\em (\bibinfo{series}{ACM SIGCOMM '23})}. \bibinfo{publisher}{Association
  for Computing Machinery}, \bibinfo{address}{New York, NY, USA},
  \bibinfo{pages}{932–949}.
\newblock
\showISBNx{9798400702365}
\showDOI{%
\url{https://doi.org/10.1145/3603269.3604870}}


\bibitem[\protect\citeauthoryear{McMillan and Zuck}{McMillan and Zuck}{2019a}]%
        {quic2}
\bibfield{author}{\bibinfo{person}{Kenneth~L McMillan} {and}
  \bibinfo{person}{Lenore~D Zuck}.} \bibinfo{year}{2019}\natexlab{a}.
\newblock \showarticletitle{Compositional testing of internet protocols}. In
  \bibinfo{booktitle}{{\em 2019 IEEE Cybersecurity Development (SecDev)}}.
  IEEE, \bibinfo{pages}{161--174}.
\newblock


\bibitem[\protect\citeauthoryear{McMillan and Zuck}{McMillan and Zuck}{2019b}]%
        {quic}
\bibfield{author}{\bibinfo{person}{Kenneth~L. McMillan} {and}
  \bibinfo{person}{Lenore~D. Zuck}.} \bibinfo{year}{2019}\natexlab{b}.
\newblock \showarticletitle{Formal Specification and Testing of QUIC}. In
  \bibinfo{booktitle}{{\em Proceedings of the ACM Special Interest Group on
  Data Communication}} {\em (\bibinfo{series}{SIGCOMM '19})}.
  \bibinfo{publisher}{Association for Computing Machinery},
  \bibinfo{address}{New York, NY, USA}, \bibinfo{pages}{227–240}.
\newblock
\showISBNx{9781450359566}
\showDOI{%
\url{https://doi.org/10.1145/3341302.3342087}}


\bibitem[\protect\citeauthoryear{Mehrotra}{Mehrotra}{2023}]%
        {outage-akamai}
\bibfield{author}{\bibinfo{person}{Shikhar Mehrotra}.}
  \bibinfo{year}{2023}\natexlab{}.
\newblock \bibinfo{title}{What Led To Internet Outage That Took Down Some Major
  Websites On July 22?}
\newblock
  \bibinfo{howpublished}{\url{https://www.republicworld.com/technology-news/other-tech-news/what-led-to-internet-outage-that-took-down-some-major-websites-on-july-22-check-out-why.html}}.
    (\bibinfo{year}{2023}).
\newblock
\newblock
\shownote{Accessed: 2023-9-29.}


\bibitem[\protect\citeauthoryear{Meng, Mirchev, B{\"o}hme, and
  Roychoudhury}{Meng et~al\mbox{.}}{2024}]%
        {chatafl}
\bibfield{author}{\bibinfo{person}{Ruijie Meng}, \bibinfo{person}{Martin
  Mirchev}, \bibinfo{person}{Marcel B{\"o}hme}, {and} \bibinfo{person}{Abhik
  Roychoudhury}.} \bibinfo{year}{2024}\natexlab{}.
\newblock \showarticletitle{Large language model guided protocol fuzzing}.
  NDSS.
\newblock


\bibitem[\protect\citeauthoryear{Merkel}{Merkel}{2014}]%
        {Docker}
\bibfield{author}{\bibinfo{person}{Dirk Merkel}.}
  \bibinfo{year}{2014}\natexlab{}.
\newblock \showarticletitle{Docker: Lightweight Linux Containers for Consistent
  Development and Deployment}.
\newblock \bibinfo{journal}{{\em Linux J.\/}} \bibinfo{volume}{2014},
  \bibinfo{number}{239} (\bibinfo{date}{March} \bibinfo{year}{2014}),
  \bibinfo{pages}{2}.
\newblock
\showISSN{1075-3583}


\bibitem[\protect\citeauthoryear{Microsoft}{Microsoft}{2023}]%
        {azure-openai}
\bibfield{author}{\bibinfo{person}{Microsoft}.}
  \bibinfo{year}{2023}\natexlab{}.
\newblock \bibinfo{title}{Azure OpenAI Service}.
\newblock
  \bibinfo{howpublished}{\\\url{https://azure.microsoft.com/en-us/products/ai-services/openai-service}}.
    (\bibinfo{year}{2023}).
\newblock


\bibitem[\protect\citeauthoryear{Mondal, Tang, Beckett, Millstein, and
  Varghese}{Mondal et~al\mbox{.}}{2023}]%
        {hotnets}
\bibfield{author}{\bibinfo{person}{Rajdeep Mondal}, \bibinfo{person}{Alan
  Tang}, \bibinfo{person}{Ryan Beckett}, \bibinfo{person}{Todd Millstein},
  {and} \bibinfo{person}{George Varghese}.} \bibinfo{year}{2023}\natexlab{}.
\newblock \showarticletitle{What do LLMs need to Synthesize Correct Router
  Configurations?}. In \bibinfo{booktitle}{{\em Proceedings of the 22nd ACM
  Workshop on Hot Topics in Networks}} {\em (\bibinfo{series}{HotNets '23})}.
  \bibinfo{publisher}{Association for Computing Machinery},
  \bibinfo{address}{New York, NY, USA}, \bibinfo{pages}{189–195}.
\newblock
\showISBNx{9798400704154}
\showDOI{%
\url{https://doi.org/10.1145/3626111.3628194}}


\bibitem[\protect\citeauthoryear{Natella}{Natella}{2022}]%
        {aflstate}
\bibfield{author}{\bibinfo{person}{Roberto Natella}.}
  \bibinfo{year}{2022}\natexlab{}.
\newblock \showarticletitle{Stateafl: Greybox fuzzing for stateful network
  servers}.
\newblock \bibinfo{journal}{{\em Empirical Software Engineering\/}}
  \bibinfo{volume}{27}, \bibinfo{number}{7} (\bibinfo{year}{2022}),
  \bibinfo{pages}{191}.
\newblock


\bibitem[\protect\citeauthoryear{Newman}{Newman}{2023}]%
        {bug-ipnet}
\bibfield{author}{\bibinfo{person}{Lily~Hay Newman}.}
  \bibinfo{year}{2023}\natexlab{}.
\newblock \bibinfo{title}{Decades-Old Code Is Putting Millions of Critical
  Devices at Risk}.
\newblock
  \bibinfo{howpublished}{\url{https://cybersecuritynews.com/bgp-error-handling-flaw/}}.
    (\bibinfo{year}{2023}).
\newblock
\newblock
\shownote{Accessed: 2023-9-29.}


\bibitem[\protect\citeauthoryear{Pacheco and Ernst}{Pacheco and Ernst}{2007}]%
        {randoop}
\bibfield{author}{\bibinfo{person}{Carlos Pacheco} {and}
  \bibinfo{person}{Michael~D Ernst}.} \bibinfo{year}{2007}\natexlab{}.
\newblock \showarticletitle{Randoop: feedback-directed random testing for
  Java}. In \bibinfo{booktitle}{{\em Companion to the 22nd ACM SIGPLAN
  conference on Object-oriented programming systems and applications
  companion}}. \bibinfo{pages}{815--816}.
\newblock


\bibitem[\protect\citeauthoryear{Pedrosa, Fogel, Kothari, Govindan, Mahajan,
  and Millstein}{Pedrosa et~al\mbox{.}}{2015}]%
        {protocol-symbex}
\bibfield{author}{\bibinfo{person}{Luis Pedrosa}, \bibinfo{person}{Ari Fogel},
  \bibinfo{person}{Nupur Kothari}, \bibinfo{person}{Ramesh Govindan},
  \bibinfo{person}{Ratul Mahajan}, {and} \bibinfo{person}{Todd Millstein}.}
  \bibinfo{year}{2015}\natexlab{}.
\newblock \showarticletitle{Analyzing protocol implementations for
  interoperability}. In \bibinfo{booktitle}{{\em 12th USENIX Symposium on
  Networked Systems Design and Implementation (NSDI 15)}}.
  \bibinfo{pages}{485--498}.
\newblock


\bibitem[\protect\citeauthoryear{Pfeil}{Pfeil}{2023}]%
        {bug-pop3}
\bibfield{author}{\bibinfo{person}{Ken Pfeil}.}
  \bibinfo{year}{2023}\natexlab{}.
\newblock \bibinfo{title}{Buffer Overflow in Digital Mapping System's POP3
  Server}.
\newblock
  \bibinfo{howpublished}{\url{https://www.itprotoday.com/email-and-calendaring/buffer-overflow-digital-mapping-systems-pop3-server}}.
    (\bibinfo{year}{2023}).
\newblock
\newblock
\shownote{Accessed: 2023-9-29.}


\bibitem[\protect\citeauthoryear{Pham, B{\"o}hme, and Roychoudhury}{Pham
  et~al\mbox{.}}{2020}]%
        {aflnet}
\bibfield{author}{\bibinfo{person}{Van-Thuan Pham}, \bibinfo{person}{Marcel
  B{\"o}hme}, {and} \bibinfo{person}{Abhik Roychoudhury}.}
  \bibinfo{year}{2020}\natexlab{}.
\newblock \showarticletitle{AFLNet: a greybox fuzzer for network protocols}. In
  \bibinfo{booktitle}{{\em 2020 IEEE 13th International Conference on Software
  Testing, Validation and Verification (ICST)}}. IEEE,
  \bibinfo{pages}{460--465}.
\newblock


\bibitem[\protect\citeauthoryear{Pham, B{\"o}hme, Santosa, C{\u{a}}ciulescu,
  and Roychoudhury}{Pham et~al\mbox{.}}{2019}]%
        {aflsmart}
\bibfield{author}{\bibinfo{person}{Van-Thuan Pham}, \bibinfo{person}{Marcel
  B{\"o}hme}, \bibinfo{person}{Andrew~E Santosa},
  \bibinfo{person}{Alexandru~R{\u{a}}zvan C{\u{a}}ciulescu}, {and}
  \bibinfo{person}{Abhik Roychoudhury}.} \bibinfo{year}{2019}\natexlab{}.
\newblock \showarticletitle{Smart greybox fuzzing}.
\newblock \bibinfo{journal}{{\em IEEE Transactions on Software Engineering\/}}
  \bibinfo{volume}{47}, \bibinfo{number}{9} (\bibinfo{year}{2019}),
  \bibinfo{pages}{1980--1997}.
\newblock


\bibitem[\protect\citeauthoryear{Romijn}{Romijn}{2023}]%
        {outage-ripe}
\bibfield{author}{\bibinfo{person}{Erik Romijn}.}
  \bibinfo{year}{2023}\natexlab{}.
\newblock \bibinfo{title}{RIPE NCC and Duke University BGP Experiment}.
\newblock
  \bibinfo{howpublished}{\url{https://labs.ripe.net/author/erik/ripe-ncc-and-duke-university-bgp-experiment/}}.
    (\bibinfo{year}{2023}).
\newblock
\newblock
\shownote{Accessed: 2023-9-29.}


\bibitem[\protect\citeauthoryear{Runciman, Naylor, and Lindblad}{Runciman
  et~al\mbox{.}}{2008}]%
        {smallcheck}
\bibfield{author}{\bibinfo{person}{Colin Runciman}, \bibinfo{person}{Matthew
  Naylor}, {and} \bibinfo{person}{Fredrik Lindblad}.}
  \bibinfo{year}{2008}\natexlab{}.
\newblock \showarticletitle{Smallcheck and lazy smallcheck: automatic
  exhaustive testing for small values}.
\newblock \bibinfo{journal}{{\em Acm sigplan notices\/}} \bibinfo{volume}{44},
  \bibinfo{number}{2} (\bibinfo{year}{2008}), \bibinfo{pages}{37--48}.
\newblock


\bibitem[\protect\citeauthoryear{Sch{\"a}fer, Nadi, Eghbali, and
  Tip}{Sch{\"a}fer et~al\mbox{.}}{2023}]%
        {llm-unit-tests}
\bibfield{author}{\bibinfo{person}{Max Sch{\"a}fer}, \bibinfo{person}{Sarah
  Nadi}, \bibinfo{person}{Aryaz Eghbali}, {and} \bibinfo{person}{Frank Tip}.}
  \bibinfo{year}{2023}\natexlab{}.
\newblock \showarticletitle{Adaptive test generation using a large language
  model}.
\newblock \bibinfo{journal}{{\em arXiv preprint arXiv:2302.06527\/}}
  (\bibinfo{year}{2023}).
\newblock


\bibitem[\protect\citeauthoryear{Sharma and Yegneswaran}{Sharma and
  Yegneswaran}{2023}]%
        {prosper}
\bibfield{author}{\bibinfo{person}{Prakhar Sharma} {and} \bibinfo{person}{Vinod
  Yegneswaran}.} \bibinfo{year}{2023}\natexlab{}.
\newblock \showarticletitle{PROSPER: Extracting Protocol Specifications Using
  Large Language Models}. In \bibinfo{booktitle}{{\em Proceedings of the 22nd
  ACM Workshop on Hot Topics in Networks}}. \bibinfo{pages}{41--47}.
\newblock


\bibitem[\protect\citeauthoryear{Siddiq, Santos, Tanvir, Ulfat, Rifat, and
  Lopes}{Siddiq et~al\mbox{.}}{2023}]%
        {llm-unit-tests3}
\bibfield{author}{\bibinfo{person}{Mohammed~Latif Siddiq},
  \bibinfo{person}{Joanna Santos}, \bibinfo{person}{Ridwanul~Hasan Tanvir},
  \bibinfo{person}{Noshin Ulfat}, \bibinfo{person}{Fahmid~Al Rifat}, {and}
  \bibinfo{person}{Vinicius~Carvalho Lopes}.} \bibinfo{year}{2023}\natexlab{}.
\newblock \showarticletitle{Exploring the Effectiveness of Large Language
  Models in Generating Unit Tests}.
\newblock \bibinfo{journal}{{\em arXiv preprint arXiv:2305.00418\/}}
  (\bibinfo{year}{2023}).
\newblock


\bibitem[\protect\citeauthoryear{Singha, Mondal, Beckett, Kakarla, Millstein,
  and Varghese}{Singha et~al\mbox{.}}{2024}]%
        {messi}
\bibfield{author}{\bibinfo{person}{Rathin Singha}, \bibinfo{person}{Rajdeep
  Mondal}, \bibinfo{person}{Ryan Beckett}, \bibinfo{person}{Siva Kesava~Reddy
  Kakarla}, \bibinfo{person}{Todd Millstein}, {and} \bibinfo{person}{George
  Varghese}.} \bibinfo{year}{2024}\natexlab{}.
\newblock \showarticletitle{$\{$MESSI$\}$: Behavioral Testing of $\{$BGP$\}$
  Implementations}. In \bibinfo{booktitle}{{\em 21st USENIX Symposium on
  Networked Systems Design and Implementation (NSDI 24)}}.
  \bibinfo{pages}{1009--1023}.
\newblock


\bibitem[\protect\citeauthoryear{Sun, Sheng, Padon, and Barrett}{Sun
  et~al\mbox{.}}{2024}]%
        {clover}
\bibfield{author}{\bibinfo{person}{Chuyue Sun}, \bibinfo{person}{Ying Sheng},
  \bibinfo{person}{Oded Padon}, {and} \bibinfo{person}{Clark Barrett}.}
  \bibinfo{year}{2024}\natexlab{}.
\newblock \showarticletitle{Clover: Closed-Loop Verifiable Code Generation}. In
  \bibinfo{booktitle}{{\em AI Verification: First International Symposium, SAIV
  2024, Montreal, QC, Canada, July 22–23, 2024, Proceedings}}.
  \bibinfo{publisher}{Springer-Verlag}, \bibinfo{address}{Berlin, Heidelberg},
  \bibinfo{pages}{134–155}.
\newblock
\showISBNx{978-3-031-65111-3}
\showDOI{%
\url{https://doi.org/10.1007/978-3-031-65112-0_7}}


\bibitem[\protect\citeauthoryear{Tufano, Drain, Svyatkovskiy, Deng, and
  Sundaresan}{Tufano et~al\mbox{.}}{2020}]%
        {llm-unit-tests2}
\bibfield{author}{\bibinfo{person}{Michele Tufano}, \bibinfo{person}{Dawn
  Drain}, \bibinfo{person}{Alexey Svyatkovskiy}, \bibinfo{person}{Shao~Kun
  Deng}, {and} \bibinfo{person}{Neel Sundaresan}.}
  \bibinfo{year}{2020}\natexlab{}.
\newblock \showarticletitle{Unit test case generation with transformers and
  focal context}.
\newblock \bibinfo{journal}{{\em arXiv preprint arXiv:2009.05617\/}}
  (\bibinfo{year}{2020}).
\newblock


\bibitem[\protect\citeauthoryear{Utting, Pretschner, and Legeard}{Utting
  et~al\mbox{.}}{2012}]%
        {mbt-overview}
\bibfield{author}{\bibinfo{person}{Mark Utting}, \bibinfo{person}{Alexander
  Pretschner}, {and} \bibinfo{person}{Bruno Legeard}.}
  \bibinfo{year}{2012}\natexlab{}.
\newblock \showarticletitle{A taxonomy of model-based testing approaches}.
\newblock \bibinfo{journal}{{\em Software testing, verification and
  reliability\/}} \bibinfo{volume}{22}, \bibinfo{number}{5}
  (\bibinfo{year}{2012}), \bibinfo{pages}{297--312}.
\newblock


\bibitem[\protect\citeauthoryear{Vikram, Lemieux, and Padhye}{Vikram
  et~al\mbox{.}}{2023}]%
        {llm-pbt}
\bibfield{author}{\bibinfo{person}{Vasudev Vikram}, \bibinfo{person}{Caroline
  Lemieux}, {and} \bibinfo{person}{Rohan Padhye}.}
  \bibinfo{year}{2023}\natexlab{}.
\newblock \showarticletitle{Can Large Language Models Write Good Property-Based
  Tests?}
\newblock \bibinfo{journal}{{\em arXiv preprint arXiv:2307.04346\/}}
  (\bibinfo{year}{2023}).
\newblock


\bibitem[\protect\citeauthoryear{Wang, Chen, Wei, and Liu}{Wang
  et~al\mbox{.}}{2019}]%
        {superion}
\bibfield{author}{\bibinfo{person}{Junjie Wang}, \bibinfo{person}{Bihuan Chen},
  \bibinfo{person}{Lei Wei}, {and} \bibinfo{person}{Yang Liu}.}
  \bibinfo{year}{2019}\natexlab{}.
\newblock \showarticletitle{Superion: Grammar-aware greybox fuzzing}. In
  \bibinfo{booktitle}{{\em 2019 IEEE/ACM 41st International Conference on
  Software Engineering (ICSE)}}. IEEE, \bibinfo{pages}{724--735}.
\newblock


\bibitem[\protect\citeauthoryear{Wang, Huang, Chen, Liu, Wang, and Wang}{Wang
  et~al\mbox{.}}{2023}]%
        {llm-testing-survey}
\bibfield{author}{\bibinfo{person}{Junjie Wang}, \bibinfo{person}{Yuchao
  Huang}, \bibinfo{person}{Chunyang Chen}, \bibinfo{person}{Zhe Liu},
  \bibinfo{person}{Song Wang}, {and} \bibinfo{person}{Qing Wang}.}
  \bibinfo{year}{2023}\natexlab{}.
\newblock \showarticletitle{Software testing with large language model: Survey,
  landscape, and vision}.
\newblock \bibinfo{journal}{{\em arXiv preprint arXiv:2307.07221\/}}
  (\bibinfo{year}{2023}).
\newblock


\bibitem[\protect\citeauthoryear{Wikipedia}{Wikipedia}{2023}]%
        {outage-rogers}
\bibfield{author}{\bibinfo{person}{Wikipedia}.}
  \bibinfo{year}{2023}\natexlab{}.
\newblock \bibinfo{title}{2022 Rogers Communications outage}.
\newblock
  \bibinfo{howpublished}{\url{https://en.wikipedia.org/wiki/2022_Rogers_Communications_outage}}.
    (\bibinfo{year}{2023}).
\newblock
\newblock
\shownote{Accessed: 2023-9-29.}


\bibitem[\protect\citeauthoryear{Woo, Cha, Gottlieb, and Brumley}{Woo
  et~al\mbox{.}}{2013}]%
        {blackbox-sched}
\bibfield{author}{\bibinfo{person}{Maverick Woo}, \bibinfo{person}{Sang~Kil
  Cha}, \bibinfo{person}{Samantha Gottlieb}, {and} \bibinfo{person}{David
  Brumley}.} \bibinfo{year}{2013}\natexlab{}.
\newblock \showarticletitle{Scheduling black-box mutational fuzzing}. In
  \bibinfo{booktitle}{{\em Proceedings of the 2013 ACM SIGSAC conference on
  Computer \& communications security}}. \bibinfo{pages}{511--522}.
\newblock


\bibitem[\protect\citeauthoryear{Xia, Paltenghi, Tian, Pradel, and Zhang}{Xia
  et~al\mbox{.}}{2023}]%
        {llm-fuzzing}
\bibfield{author}{\bibinfo{person}{Chunqiu~Steven Xia}, \bibinfo{person}{Matteo
  Paltenghi}, \bibinfo{person}{Jia~Le Tian}, \bibinfo{person}{Michael Pradel},
  {and} \bibinfo{person}{Lingming Zhang}.} \bibinfo{year}{2023}\natexlab{}.
\newblock \showarticletitle{Universal Fuzzing via Large Language Models}.
\newblock \bibinfo{journal}{{\em arXiv preprint arXiv:2308.04748\/}}
  (\bibinfo{year}{2023}).
\newblock


\bibitem[\protect\citeauthoryear{Ye, Tang, Tan, Huang, Fang, Sun, Bian, Wang,
  and Wang}{Ye et~al\mbox{.}}{2021}]%
        {llm-conformance-tests}
\bibfield{author}{\bibinfo{person}{Guixin Ye}, \bibinfo{person}{Zhanyong Tang},
  \bibinfo{person}{Shin~Hwei Tan}, \bibinfo{person}{Songfang Huang},
  \bibinfo{person}{Dingyi Fang}, \bibinfo{person}{Xiaoyang Sun},
  \bibinfo{person}{Lizhong Bian}, \bibinfo{person}{Haibo Wang}, {and}
  \bibinfo{person}{Zheng Wang}.} \bibinfo{year}{2021}\natexlab{}.
\newblock \showarticletitle{Automated conformance testing for javascript
  engines via deep compiler fuzzing}. In \bibinfo{booktitle}{{\em Proceedings
  of the 42nd ACM SIGPLAN international conference on programming language
  design and implementation}}. \bibinfo{pages}{435--450}.
\newblock


\bibitem[\protect\citeauthoryear{Yen, L{\'e}vai, Ye, Ren, Govindan, and
  Raghavan}{Yen et~al\mbox{.}}{2021}]%
        {nlp-testing}
\bibfield{author}{\bibinfo{person}{Jane Yen}, \bibinfo{person}{Tam{\'a}s
  L{\'e}vai}, \bibinfo{person}{Qinyuan Ye}, \bibinfo{person}{Xiang Ren},
  \bibinfo{person}{Ramesh Govindan}, {and} \bibinfo{person}{Barath Raghavan}.}
  \bibinfo{year}{2021}\natexlab{}.
\newblock \showarticletitle{Semi-automated protocol disambiguation and code
  generation}. In \bibinfo{booktitle}{{\em Proceedings of the 2021 ACM SIGCOMM
  2021 Conference}}. \bibinfo{pages}{272--286}.
\newblock


\bibitem[\protect\citeauthoryear{Yuan, Lou, Liu, Ding, Wang, Chen, and
  Peng}{Yuan et~al\mbox{.}}{2023}]%
        {llm-unit-tests4}
\bibfield{author}{\bibinfo{person}{Zhiqiang Yuan}, \bibinfo{person}{Yiling
  Lou}, \bibinfo{person}{Mingwei Liu}, \bibinfo{person}{Shiji Ding},
  \bibinfo{person}{Kaixin Wang}, \bibinfo{person}{Yixuan Chen}, {and}
  \bibinfo{person}{Xin Peng}.} \bibinfo{year}{2023}\natexlab{}.
\newblock \showarticletitle{No More Manual Tests? Evaluating and Improving
  ChatGPT for Unit Test Generation}.
\newblock \bibinfo{journal}{{\em arXiv preprint arXiv:2305.04207\/}}
  (\bibinfo{year}{2023}).
\newblock


\bibitem[\protect\citeauthoryear{Zare and Community}{Zare and
  Community}{2023}]%
        {technitium}
\bibfield{author}{\bibinfo{person}{Shreyas Zare} {and}
  \bibinfo{person}{Community}.} \bibinfo{year}{2023}\natexlab{}.
\newblock \bibinfo{title}{Technitium DNS Server}.
\newblock \bibinfo{howpublished}{\\\url{https://technitium.com/dns/}}.
  (\bibinfo{year}{2023}).
\newblock
\newblock
\shownote{\\Eywa commit:
  \url{https://github.com/TechnitiumSoftware/DnsServer/tree/d4352680b3f14fa2884fc8b7fa9c7772379bbc61}.}


\end{thebibliography}
\clearpage

\appendix

\section{Regex Implementation}
\label{sec:appendix-regex}

We show \tool's minimal regular expression implementation in \cref{fig:appendix-regex}. The implementation supports regular expression range, union, sequence, and iteration (star) constructs, defined by simple \ccode{Regex} type. To preserve branches for symbolic execution, rather than build a finite automaton, the matching logic uses a notion of regular expression continuation \ccode{RegexCont}, which is a linked list of regular expressions that must be matched after the current regular expression. Initially, this list is empty but gets added to any time there is a regular expression sequence (\ccode{regex->op == SEQ}) case. Continuations allow for a simple recursive implementation, and work well with symbolic execution since the regular expression itself is always concrete. Path constraints come from matches against the symbolic text \ccode{*text} such as the comparison with \ccode{'\0'} and the case for \ccode{regex->op == RANGE}. For the \ccode{match} function, we pass in a ``null'' continuation to begin with.

\begin{lrbox}{\codei}
\begin{lstlisting}[style=CStyle]
typedef enum { OR, SEQ, STAR, RANGE } RegexOp;
typedef struct Regex Regex;
struct Regex {
    RegexOp op; int clo; int chi;
    Regex* left;Regex* right;
};
typedef struct RegexCont RegexCont;
struct RegexCont {
    Regex* regex;
    RegexCont* next;
}
// Matching logic for a regular expression with a
// continuation.
static int match(Regex* regex, RegexCont* cont, 
char *text) {
  if (regex == NULL) {return *text == '\0';}
  if (regex->op == OR) {
    return match(regex->left, cont, text) ||
    match(regex->right, cont, text);
  }
  if (regex->op == SEQ) {
    RegexCont c;
    c.next = cont; c.regex = regex->right;
    return match(regex->left, &c, text);
  }
  if (regex->op == STAR) {
    Regex r; r.op = SEQ; r.left = regex->left;
    r.right = regex;
    return match(cont->regex, cont->next, text) ||
    (*text != '\0' && match(&r, cont, text));
  }
  if (regex->op == RANGE) {
    char c = *text++;
    return c != '\0' && c >= regex->clo && 
    c <= regex->chi && 
    match(cont->regex, cont->next, text);
  }
  return 0;
}
// Matching logic for a regular expression.
static int match(Regex* regex, char *text) {
  RegexCont cont;
  cont.next = NULL; cont.regex = NULL;
  return match(regex, &cont, text);
}
\end{lstlisting}
\end{lrbox}

\begin{figure}[t!]
\centering
\begin{tikzpicture}
    \node[inner sep=2pt] (code) at (0,0) {
      \begin{tikzpicture}
        \node[inner sep=5pt,rounded corners=.1cm,draw=gray,fill=gray!0,thick] (impl) at (0,0) { \usebox\codei };
      \end{tikzpicture}};
\end{tikzpicture}
\vspace{-1em}
\caption{\tool's minimal regular expression implementation that is amenable to symbolic execution.}
\label{fig:appendix-regex}
\end{figure}

For each RegexModule, \tool converts them to functions with C expressions that evaluate to boolean values. The translation is straightforward -- for instance, a Python constraint such as
\pcode{eywa.RegexModule("[a-z]*", a1)}
will generate the C code:
\begin{lstlisting}[style=CStyle]
Regex r1; r1.clo = 'a'; r1.chi = 'z';
Regex r2; r2.op = STAR; r2.left = r1;
klee_assume(match(&r2, &a1));
\end{lstlisting}
\tool then calls a custom \ccode{match} function to check if the string matches the regular expression. The \ccode{match} function is a minimal regular expression matching implementation written by hand and is amenable to symbolic execution.
Klee will encode the
regex match condition and other conditions symbolically
and solve them as part of its path exploration.

\section{Hyperparameter selection}
\label{sec:appendix-hyperparameter}

\pgfplotstableread[col sep=comma]{
K, 0.0, 0.2, 0.4, 0.6, 0.8, 1.0
1, 249.00, 249.00, 245.44, 249.00, 263.11, 266.33
2, 249.33, 249.44, 251.44, 256.11, 270.50, 296.88
3, 251.44, 249.44, 253.89, 259.38, 283.44, 305.38
4, 251.44, 249.44, 261.11, 265.56, 284.88, 335.56
5, 251.44, 249.44, 266.00, 264.88, 287.89, 351.78
6, 251.44, 249.44, 266.11, 276.88, 289.44, 357.12
7, 251.44, 251.89, 266.33, 280.75, 300.11, 360.89
8, 251.44, 254.22, 285.56, 277.33, 304.56, 380.11
9, 251.44, 254.33, 285.56, 292.89, 306.56, 412.29
10, 255.00, 256.56, 290.12, 292.89, 330.00, 400.44
11, 255.00, 256.56, 290.00, 292.89, 349.56, 410.56
12, 255.00, 273.33, 292.22, 295.67, 354.00, 435.12
}\dnamedatatable

\pgfplotstableread[col sep=comma]{
K, 0.0, 0.2, 0.4, 0.6, 0.8, 1.0
1, 242.89, 250.44, 262.67, 258.56, 268.78, 286.56
2, 284.00, 323.22, 307.22, 321.11, 344.25, 372.44
3, 322.56, 348.89, 334.11, 365.44, 387.88, 411.78
4, 335.89, 377.22, 368.33, 417.12, 412.78, 460.12
5, 344.78, 384.56, 386.89, 432.00, 442.56, 482.11
6, 356.67, 389.11, 404.44, 456.57, 453.56, 499.62
7, 368.78, 397.89, 414.67, 463.78, 457.75, 503.78
8, 380.56, 416.78, 424.22, 470.67, 473.89, 523.11
9, 392.56, 424.67, 443.25, 491.89, 485.89, 536.56
10, 401.22, 437.22, 468.00, 492.33, 490.38, 538.56
11, 405.11, 444.44, 479.89, 514.22, 503.67, 550.50
12, 405.56, 450.33, 490.89, 526.56, 512.78, 550.89
}\ipdatatable

\pgfplotstableread[col sep=comma]{
K, 0.0, 0.2, 0.4, 0.6, 0.8, 1.0
1, 208.33, 214.89, 245.78, 230.00, 226.89, 250.00
2, 229.89, 254.67, 303.56, 280.00, 266.78, 326.75
3, 259.00, 265.78, 330.56, 304.67, 309.00, 370.44
4, 280.11, 268.44, 343.67, 327.44, 343.67, 416.00
5, 298.11, 315.00, 358.11, 345.78, 356.56, 437.67
6, 317.56, 319.78, 361.44, 366.89, 367.78, 440.67
7, 318.22, 320.67, 371.78, 380.11, 380.25, 461.67
8, 323.89, 331.22, 384.67, 384.00, 397.33, 478.00
9, 323.89, 342.33, 387.44, 397.56, 409.22, 492.62
10, 336.11, 346.89, 394.44, 404.22, 413.22, 491.56
11, 338.00, 348.78, 403.22, 408.56, 416.89, 496.44
12, 338.00, 368.11, 417.56, 419.67, 432.56, 502.22
}\wildcarddatatable

\pgfplotstableread[col sep=comma]{
K, 0.2, 0.4, 0.6, 0.8, 1.0
1, 281.50, 272.20, 275.67, 270.11, 250.40
2, 349.20, 356.40, 317.30, 330.67, 330.89
3, 400.60, 385.60, 380.90, 371.12, 373.80
4, 428.20, 424.90, 435.20, 390.44, 407.30
5, 447.50, 445.50, 448.90, 435.22, 461.10
6, 451.60, 453.00, 480.75, 458.44, 499.22
7, 462.60, 470.60, 493.78, 469.78, 507.00
8, 476.00, 478.70, 511.10, 494.33, 511.20
9, 478.00, 485.80, 521.89, 499.33, 531.00
10, 484.60, 496.40, 542.44, 515.67, 540.70
11, 490.10, 502.70, 541.20, 538.67, 563.80
12, 492.60, 510.70, 555.40, 546.62, 576.20
}\datatable

\begin{figure*}
    \centering
    \begin{subfigure}[b]{\textwidth}
        \centering
        \pgfplotslegendfromname{commonlegend}
    \end{subfigure}
    
    \begin{subfigure}[b]{0.45\textwidth}
     \begin{tikzpicture}
        \begin{axis}[
            xlabel={$k$},
            ylabel={unique tests},
            legend pos=south east,
            grid=major,
            ymin=0,
            ymax=650,
            width=\columnwidth,
            height=0.64\columnwidth,
            legend entries={$\tau = 0.2$, $\tau = 0.4$, $\tau = 0.6$, $\tau = 0.8$, $\tau = 1.0$},
            legend to name=commonlegend,  
            legend columns=5,  
            legend style={
                draw=none,
                column sep=3ex,
                inner xsep=0pt,
                nodes={inner sep=0pt, text depth=0.15em},
                /tikz/every odd column/.append style={column sep=3pt} 
            }, 
        ]
        \addplot table[x=K, y=0.2] {\dnamedatatable};
        \addplot table[x=K, y=0.4] {\dnamedatatable};
        \addplot table[x=K, y=0.6] {\dnamedatatable};
        \addplot table[x=K, y=0.8] {\dnamedatatable};
        \addplot table[x=K, y=1.0] {\dnamedatatable};
        \end{axis}
        \end{tikzpicture}
        \caption{\textsc{dname}}
        \vspace{1em}
    \end{subfigure}
    \begin{subfigure}[b]{0.45\textwidth}
    \begin{tikzpicture}
    \begin{axis}[
        xlabel={$k$},
        ylabel={unique tests},
        legend pos=south east,
        grid=major,
        ymin=0,
        ymax=650,
        width=\columnwidth,
        height=0.64\columnwidth,
    ]
    
    \addplot table[x=K, y=0.2] {\ipdatatable};
    \addplot table[x=K, y=0.4] {\ipdatatable};
    \addplot table[x=K, y=0.6] {\ipdatatable};
    \addplot table[x=K, y=0.8] {\ipdatatable};
    \addplot table[x=K, y=1.0] {\ipdatatable};
    \end{axis}
    \end{tikzpicture}
    \caption{\textsc{ipv4}}
    \vspace{1em}
     \end{subfigure}

    \begin{subfigure}[b]{\textwidth}
        \centering
        \pgfplotslegendfromname{commonlegend}
    \end{subfigure}
    
    \begin{subfigure}[t]{0.45\textwidth}
    \begin{tikzpicture}
    \begin{axis}[
        xlabel={$k$},
        ylabel={unique tests},
        legend pos=south east,
        grid=major,
        ymin=0,
        ymax=650,
        width=\columnwidth,
        height=0.64\columnwidth,
    ]
    
    \addplot table[x=K, y=0.2] {\wildcarddatatable};
    \addplot table[x=K, y=0.4] {\wildcarddatatable};
    \addplot table[x=K, y=0.6] {\wildcarddatatable};
    \addplot table[x=K, y=0.8] {\wildcarddatatable};
    \addplot table[x=K, y=1.0] {\wildcarddatatable};
    \end{axis}
    \end{tikzpicture}
    \caption{\textsc{wildcard}},
     \end{subfigure}
     \begin{subfigure}[t]{0.45\textwidth}
        \begin{tikzpicture}
        \begin{axis}[
            xlabel={$k$},
            ylabel={unique tests},
            legend pos=south east,
            grid=major,
            ymin=0,
            ymax=650,
            width=\columnwidth,
            height=0.64\columnwidth,
        ]
        
        \addplot table[x=K, y=0.2] {\datatable};
        \addplot table[x=K, y=0.4] {\datatable};
        \addplot table[x=K, y=0.6] {\datatable};
        \addplot table[x=K, y=0.8] {\datatable};
        \addplot table[x=K, y=1.0] {\datatable};
        \end{axis}
        \end{tikzpicture}
        \caption{\textsc{cname}}
        \end{subfigure}
    \vspace{-\baselineskip}    
    \caption{Hyperparameter analysis for DNS models. The plots show how the total number of unique tests vary as we increase the number of runs for different models.}
    \label{fig:dns-hyperparameters}
\end{figure*}

To understand how the temperature $\tau$ and the number of attempts $k$ affect test case generation, we took the \textsc{cname} DNS model and varied counted the number of tests generated at for each $k$ averaged over 10 runs. We plot this count vs. $k$ in \cref{fig:dns-hyperparameters} for values of $\tau$ ranging from 0.2 to 1.0. Results for the other models were similar. We can see from the graph that there are greatly diminishing returns around $k=10$, whereas $\tau$ appears to have less effect on the test generation for values above 0. For this reason, we selected $k=10$ and $\tau=0.6$ in our experiments, which we believe represents a reasonable trade off between efficiency and test coverage.
\begin{lrbox}{\codej}
\begin{lstlisting}[style=PythonStyle]
# initialize the dependency graph
g = eywa.DependencyGraph()
# add the call edges to the graph
g.CallEdge(isValidPrefixList,
    [prefixLengthToSubnetMask])
g.CallEdge(isValidRoute,[prefixLengthToSubnetMask])
g.CallEdge(checkValidInputs,
    [isValidPrefixList,isValidRoute])
g.CallEdge(isMatchPrefixListEntry,
    [prefixLengthToSubnetMask])
g.CallEdge(isMatchRouteMapStanza,
    [isMatchPrefixListEntry])
# add the pipe
g.Pipe(checkValidInputs,isMatchRouteMapStanza)
final_model = g.Synthesize() # synthesize final model
\end{lstlisting}
\end{lrbox}
\begin{figure}[]
\centering
\begin{tikzpicture}
    \node[inner sep=2pt] (code) at (0,0) {
      \begin{tikzpicture}
        \node[inner sep=5pt,rounded corners=.1cm,draw=gray,fill=gray!0,thick] (impl) at (0,0) { \usebox\codej };
      \end{tikzpicture}};
\end{tikzpicture}
\vspace{-1em}
\caption{\tool code for building a dependency graph for the  BGP \textsc{rmap-pl} model in \cref{tab:eval-tests-and-results}}
\label{fig:dag}
\end{figure}

\section{Graph API}
\label{appendix:graph-api}

\begin{lrbox}{\codeg}
\begin{lstlisting}[style=CStyle]
#include <stdint.h>
...
typedef struct {uint32_t prefix; uint8_t prefixLength; ...} Route;
typedef struct {uint32_t prefix; uint8_t prefixLength; uint32_t le; uint32_t ge; bool any; bool permit;
} PrefixListEntry;
// a function that takes as input the prefix length and converts it to the corresponding unsigned 
// integer representation
//
// Parameters:
//     maskLength: The length of the prefix
// Return Value:
//     The unsinged integer representation of the prefix length
uint32_t prefixLengthToSubnetMask(uint32_t maskLength);
// A function that takes as input a prefix list entry and a BGP route advertisement. If the route advertisement 
// matches the prefix, then the function should return the value of the permit flag. In case there is no match, 
// the function should vacuously return false.
//
// Parameters:
//     route: Route to be matched
//     pfe: Prefix list entry
// Return Value:
//     True if the route matches the prefix list entry
bool isMatchPrefixListEntry(Route route, PrefixListEntry pfe) {
\end{lstlisting}
\end{lrbox}

\begin{figure*}[]
\centering
\begin{tikzpicture}
    \node[inner sep=2pt] (codea) at (0,0) {
      \begin{tikzpicture}
        \node[inner sep=5pt,rounded corners=.1cm,draw=gray,fill=gray!0,thick] (impl) at (0,0) { \usebox\codeg };
      \end{tikzpicture}};
\end{tikzpicture}
\vspace{-2em}
\caption{Prompt for the \texttt{prefixLengthToSubnetMask} to \texttt{isMatchPrefixListEntry} dependency in Figure \ref{fig:dag}.}
\label{fig:dependency-prompt}
\end{figure*}

\cref{fig:dag} shows the construction of the dependency graph for the \textsc{rmap-pl} BGP model from our experiments, which matches routes against a prefix list used in a route-map stanza. Each of the modules is first defined using the prompting technique shown in \cref{fig:overview-example}a. We then specify the module dependencies using Eywa's built-in graph API. We first initialize an empty graph and then add in all the dependencies using the \pcode{CallEdge} and \texttt{Pipe} methods. In this example, we use a \pcode{Pipe} between the functions \pcode{checkValidInputs} and \pcode{isMatchRouteMapStanza}, as the former is used to enforce constraints on the inputs to the latter. The rest of the edges are of the type \pcode{CallEdge}, as in those cases the result from one function is being directly used in another. We need to define multiple \pcode{CallEdges} here because there are hierarchical dependencies. For instance, the function \pcode{isMatchRouteMapStanza} depends on \pcode{isMatchPrefixListEntry} which in turn calls the function \pcode{prefixLengthToSubnetMask}. Also, the reader may observe that unlike the other modules, \pcode{checkValidInputs} is dependent on two modules, namely \pcode{isValidRoute} and \pcode{isValidPrefixList}. Both of them must therefore be specified in a list while using the \pcode{CallEdge} method. Following this, we can synthesize the final model using the \pcode{Synthesize} utility. For synthesizing individual modules with incoming \pcode{CallEdges}, the prompt includes a description of all the dependency functions along with their C prototypes. \cref{fig:dependency-prompt} shows the prompt generated due to the \pcode{CallEdge} from the function named \pcode{prefixLengthToSubnetMask} to the function \pcode{isMatchPrefixListEntry}, when we synthesize a model for the latter. This is crucial because it helps to make the LLM aware of all the utility functions available to the current module to make use of. Finally, Eywa inserts all the functions into the final program in the correct order according to their topological ordering.

\begin{lrbox}{\codel}
\begin{lstlisting}[style=CStyle]
Your goal is to implement the C function provided by the user.
The result should be the complete implementation of the code, including:
  1. All the import statements needed, including those provided in the input. All the imports from the input
  should be included.
  2. All the type definitions provided by the user. The type definitions should NOT be modified
  3. ONLY write code for the function that has 'implement me' written in its function body.
  4. If any additional function prototypes are provided, you can use them as helper functions. There is 
  no need to define them. You can assume they will be done later by the user.
  5. Do NOT change the provided function declarations/prototypes.
  6. Whenever you define a struct, write it in one line. Do not put newline. e.g. struct { int x; int y; }

Do NOT add a `main()` function or any examples, just implement the function.
DO NOT USE fenced code blocks, just write the code.
DO NOT USE C strtok function. Implement your own.

Example Input:
#include <stdint.h>
#include <stdbool.h>
#include <string.h>
#include <stdlib.h>
#include <klee/klee.h>
#include <stdio.h>

typedef uint32_t myint;

myint add_one(myint x) {
  // implement me
}
Example Output:
#include <stdint.h>
#include <stdbool.h>
#include <string.h>
#include <stdlib.h>
#include <klee/klee.h>
#include <stdio.h>

typedef uint32_t myint;

myint add_one(myint x) {
    return x + 1
}
\end{lstlisting}
\end{lrbox}

\begin{figure*}[b]
\centering
\begin{tikzpicture}
    \node[inner sep=2pt] (codea) at (0,0) {
      \begin{tikzpicture}
        \node[inner sep=5pt,rounded corners=.1cm,draw=gray,fill=gray!0,thick] (impl) at (0,0) { \usebox\codel };
      \end{tikzpicture}};
\end{tikzpicture}
\vspace{-1em}
\caption{The system prompt which we used for all LLM calls.}
\label{fig:system-prompt}
\end{figure*}

\section{System Prompt}
\label{appendix:system-prompt}

\cref{fig:system-prompt} shows the system prompt used for all the LLM calls for our experiments.

\end{document}